\documentclass[12pt]{article}

\pdfoutput=1

\usepackage{ifpdf}
\ifpdf
\usepackage{graphicx,color}
\usepackage{hyperref}
\else
\usepackage[dvipdfmx]{graphicx,color}
\usepackage[dvipdfmx]{hyperref}
\fi
\usepackage{amssymb,amsfonts,amsmath,cancel,cite,multirow}
\usepackage[capitalise]{cleveref}
\usepackage{slashed}
\usepackage{subcaption}
\usepackage[margin=1in]{geometry}

\newcommand{\eqs}[1]{\begin{equation}\begin{split} #1 \end{split}\end{equation}}

\begin{document}

\begin{titlepage}

\begin{flushright}
\end{flushright}

\vskip 1.35cm
\begin{center}

{\large
\textbf{
  Perturbative Unitarity of Strongly Interacting Massive Particle Models
}}
\vskip 1.2cm

Ayuki Kamada$^{a,b}$,
Shin Kobayashi$^{c}$,
and Takumi Kuwahara$^{d}$

\vskip 0.4cm

\textit{$^a$
Institute of Theoretical Physics, Faculty of Physics, University of Warsaw, ul. Pasteura 5, PL-02-093 Warsaw, Poland
}

\textit{$^b$
Kavli Institute for the Physics and Mathematics of the Universe (WPI), The University of Tokyo Institutes for Advanced Study, The University of Tokyo, Kashiwa 277-8583, Japan
}

\textit{$^c$
ICRR, University of Tokyo, Kashiwa, Chiba 277-8582, Japan
}

\textit{$^d$
Center for High Energy Physics, Peking University, Beijing 100871, China
}

\vskip 1.5cm

\begin{abstract}
  Dark pion is a promising candidate for the strongly interacting massive particle dark matter. 
  A large pion self-coupling $m_\pi/f_\pi$ tends to be required for correct relic abundance, and hence the partial-wave amplitudes can violate the perturbative unitarity even for the coupling within na\"ive perturbative regime.
  We improve the partial-wave amplitudes in order to satisfy the optical theorem. 
  We demonstrate that the improvement is relevant only for semi-relativistic pions, and thus this does not affect the self-scattering cross section at the cosmic structures.
  We also discuss the impact of the improvement of the $\pi \pi \pi \to \pi \pi$ scattering process, and we find that there is an upper bound on $m_\pi$ at which the correct relic abundance is never achieved even at large $m_\pi/f_\pi$ due to the optical theorem.
\end{abstract}

\end{center}
\end{titlepage}

%%%%%%%%%%%%%%%%%%%%%%%%%%%%%%%%%%%%%%%%%%%%%%%%%%%%%%%%%%%%%%%%%%%%%%%%%%%%%%%%%%%%%%%%
\section{Introduction}

Little is known about particle nature of dark matter (DM) even though the existence of DM has been firmly confirmed by the astrophysical observations.
Strongly interacting massive particles (SIMPs)~\cite{Hochberg:2014dra} are an interesting framework for the thermal relic of sub-GeV DM: the thermal relic DM is determined by the freeze-out of $3 \to 2$ processes.
Self-interactions of DM are generically sizable to get the correct relic abundance in this framework, and it leads to a large $2 \to 2$ self-scattering.
The small-scale structure of the Universe may indicate the sizable self-scattering of DM~\cite{Spergel:1999mh} (see Ref.~\cite{Tulin:2017ara} a review).

Dark sector is a hypothetical sector where DM resides, and has its own gauge dynamics. 
As a consequence of dark strong dynamics, the dark sector would consist of composite particles (dark hadrons) at the low-energy scale as with the SM hadrons~\cite{Gudnason:2006yj, Dietrich:2006cm, Khlopov:2007ic, Khlopov:2008ty, Foadi:2008qv, Mardon:2009gw, Kribs:2009fy, Barbieri:2010mn, Blennow:2010qp, Lewis:2011zb, Appelquist:2013ms, Hietanen:2013fya, Cline:2013zca, Appelquist:2014jch, Hietanen:2014xca, Krnjaic:2014xza, Detmold:2014qqa, Detmold:2014kba, Asano:2014wra, Brod:2014loa, Antipin:2014qva, Hardy:2014mqa, Appelquist:2015yfa, Appelquist:2015zfa, Antipin:2015xia, Hardy:2015boa, Co:2016akw, Dienes:2016vei, Ishida:2016fbp, Lonsdale:2017mzg, Berryman:2017twh,  Gresham:2017zqi, Gresham:2017cvl, Mitridate:2017oky, Gresham:2018anj, Ibe:2018juk, Braaten:2018xuw, Francis:2018xjd, Bai:2018dxf, Chu:2018faw, Hall:2019rld, Tsai:2020vpi, Asadi:2021yml, Asadi:2021pwo, Zhang:2021orr, Bottaro:2021aal, Hall:2021zsk, Asadi:2022vkc} (see Ref.~\cite{Kribs:2016cew} for a review). 
Ref.~\cite{Hochberg:2014kqa} proposed a model of the SIMP framework where the dark pion is identified as DM.
The dark pions arise as the pseudo-Nambu-Goldstone boson (pNGB) from the strong dynamics in the dark sector.
The dark pions have the $2 \to 2$ self-interaction and the $3 \to 2$ number-changing process induced by the Wess-Zumino-Witten (WZW) term~\cite{Wess:1971yu,Witten:1983tw}. 

The chiral perturbation theory ($\chi$PT) describes the interactions among dark pions.
The pion self-coupling, which is defined by the ratio of the pion mass and decay constant $m_\pi/f_\pi$, determines the size of the pionic scattering processes. 
The $2 \to 2$ self-scattering interaction arises at the leading order terms of $\mathcal{O}(m_\pi^2/f_\pi^2)$, while the $3 \to 2$ number-changing interaction via the WZW term appears as an $\mathcal{O}(m_\pi^5/f_\pi^5)$ term of $\chi$PT.
The pion self-coupling is found to be larger than unity in order to explain the relic abundance and to evade constraints on the self-scattering cross section~\cite{Hochberg:2014kqa}. 
Meanwhile, we cannot validate the perturbative expansion of $\chi$PT unless $m_\pi/f_\pi \lesssim 4\pi$.
We would not be able to ignore contributions from resonances and the higher-order terms of the chiral Lagrangian for the pion self-coupling near the na\"ive perturbative bound. 
For the consistent treatment of the chiral expansion, Ref.~\cite{Hansen:2015yaa} has discussed the impact of the higher order of the chiral Lagrangian on SIMP scenarios.

We may encounter the other bound on the self-scattering cross section of dark pions in the SIMP models even when the perturbative expansion of $\chi$PT is valid.
The unitarity of the $S$-matrix imposes constraints on the partial-wave amplitude.
The partial-wave amplitude for the $2 \to 2$ self-scattering is $T \simeq m_\pi^2/(32 \pi f_\pi^2)$ at the tree-level, and the perturbative unitarity places an upper bound $1/2 \beta$ on $\mathrm{Re} T$ with $\beta$ corresponding to velocity of pions.
The perturbative unitarity bound $m_\pi/f_\pi \lesssim 4 \sqrt{\pi}/\sqrt{\beta}$ gets weaker at the cosmic structures due to the DM velocity at maximum of $\sim 10^{-2}$, while this bound can be important for the annihilation mechanism determining the current relic abundance since dark pions are semi-relativistic.
In other words, there are two bounds on the pion self-couplings: one originates from the limitation of the perturbative expansion and another is the perturbative unitarity of the scattering processes.
In the chiral limit ($m_\pi \to 0$), the perturbative unitarity violation will be cured by resumming multiple rescattering processes, which is known as the ``self-healing'' mechanism~\cite{Aydemir:2012nz}.
In this paper, we will propose the improvement of $2 \to 2$ and $3 \to 2$ partial-wave amplitudes in a similar way to the ``self-healing'' mechanism in the non-chiral limit since we focus on the dark pion DM.

This paper is organized as follows.
We will show the chiral Lagrangian for the SIMP models in \cref{sec:chiL}.
We discuss the perturbative unitarity of $\pi \pi \to \pi \pi$ scattering cross section and thermally-averaged $\pi \pi \pi \to \pi \pi$ cross section in \cref{sec:unitarity}, and we will see these cross sections would violate the perturbative unitarity at large $m_\pi/f_\pi$. 
In \cref{sec:improvement}, we propose the improved amplitude, which satisfies the optical theorem automatically, and then we will apply the procedure to the SIMP models. 
\cref{sec:conclusion} is devoted to conclusions of our work.

%%%%%%%%%%%%%%%%%%%%%%%%%%%%%%%%%%%%%%%%%%%%%%%%%%%%%%%%%%%%%%%%%%%%%%%%%%%%%%%%%%%%%%%%
\section{Chiral Lagrangian for SIMP \label{sec:chiL}}

We discuss the SIMP model that is realized by the confining gauge dynamics of a gauge group $G_\mathrm{local}$. 
We consider $N_f$-flavor quarks with the mass below the dynamical scale of $G_\mathrm{local}$ in the ultraviolet description of this model. 
This model possesses an approximate global symmetry $G$ among quarks that is broken by the mass terms. 
It is believed that this model leads to the chiral symmetry breaking, and that chiral condensation breaks the global symmetry into the subgroup $H$. 
In the following, we consider three classes of the models: (\textit{i}) $G_\mathrm{local}= SU(N_c)$, $G = SU(N_f) \times SU(N_f)$, and $H = SU(N_f)$; (\textit{ii}) $G_\mathrm{local}= SO(N_c)$, $G = SU(N_f)$, and $H = SO(N_f)$; and (\textit{iii}) $G_\mathrm{local}= USp(N_c)$, $G = SU(N_f)$, and $H = USp(N_f)$ (with even integers $N_c$ and $N_f$ \cite{Peskin:1980gc,Preskill:1980mz,Witten:1983tx,Kosower:1984aw}).

The dark pions, which are the pNGBs of the chiral symmetry breaking, are the fundamental degrees of freedom in the low-energy effective theory of this model.
We are interested in the dark pions for realizing SIMP framework, and hence we focus on the chiral expansion with typical momentum $p^2 \simeq \mathcal{O}(m_\pi^2)$ in the non-relativistic limit. 
The coset space $G/H$ is parameterized by $N_\pi$ pNGB fields $\pi^a$~\cite{Coleman:1969sm,Callan:1969sn}, which corresponds to the broken generators $T^a$ with $a = 1, \cdots, N_\pi$.
The parameterization and the normalization of the dark pion fields are the following.
\eqs{
  \Sigma = \exp \left( \frac{2 i \Pi}{f_\pi} \right) \,, \qquad 
  \Pi \equiv \pi^a T^a \,, \qquad 
  \mathrm{Tr}(T^a T^b) = \frac12 \delta^{ab} \,,
}
and only for the case (\textit{iii}), the non-linear sigma model field $\Sigma$ is 
\eqs{
  \Sigma = \exp \left( \frac{2 i \Pi}{f_\pi} \right) J \,,
}
where $J$ is the symplectic metric that satisfies $J^T = - J \,, J^2 = - \mathbf{1}$.
Here, $f_\pi$ denotes the pion decay constant.
Under the residual symmetry $H$, the dark pion fields transform as (\textit{i}) adjoint representation, (\textit{ii}) rank-2 symmetric tensor representation, and (\textit{iii}) rank-2 antisymmetric tensor representation for each different symmetry.
The relevant Lagrangian of the dark pions is given by
\eqs{
  \mathcal{L}
   = \frac{f_\pi^2}{4} \mathrm{Tr} \left[ (D_\mu \Sigma)^\dag D^\mu \Sigma \right]
   + \frac{B f_\pi^2}{4} \mathrm{Tr} \left[ M^\dag \Sigma + M \Sigma^\dag \right]
   + \mathcal{L}_\mathrm{WZW} \,. 
}
Here, the first-two terms are the leading order (LO) terms in the $\chi$PT and the soft chiral symmetry breaking term, and the third term is the Wess-Zumino-Witten (WZW) term~\cite{Wess:1971yu,Witten:1983tw}.
We take the quark mass matrix $M$ to be invariant under $H$, and hence the pion mass is universal $m_\pi^2 = B m_q$ with $m_q$ being the quark mass.

The relevant Lagrangian contains the four-point interactions among the dark pions, which induce the self scattering of the dark pions.
We obtain the four-point interaction terms from the kinetic and mass terms by expanding $\Sigma$.
\eqs{
  \mathcal{L}
  \supset \frac{r_{abcd}}{4 f_\pi^2} \pi^a \pi^b \partial_\mu \pi^c \partial^\mu \pi^d
  + \frac{1}{24} \frac{m_\pi^2}{f_\pi^2} c_{abcd} \pi^a \pi^b \pi^c \pi^d \,,
  \label{eq:4pi}
}
where $r_{abcd}$ and $c_{abcd}$ are the coefficients defined by group-theoretical constants, which is discussed in \cref{app:group}.
We will discuss the self-scattering cross section in detail in the next section.
Meanwhile, the five-point interaction among the dark pions arises from the WZW term. 
The $\pi\pi\pi \to \pi\pi$ scattering process arising from the WZW term determines the relic abundance of the dark pions. 
The WZW term is given by%
\footnote{
  Our conventions are different from some literature.
  The normalizations of both generators and $f_\pi$ differ factor two from Refs~\cite{Witten:1983tw,Hochberg:2014kqa}, and the resultant coefficients are same as ours.
  Meanwhile, the normalization of generators differs, but the same convention is used for $f_\pi$ as ours in \cite{Katz:2020ywn}.
}
\eqs{
  \mathcal{L}_\mathrm{WZW} 
  & = \frac{2k}{15\pi^2 f_\pi^5} \epsilon^{\mu\nu\rho \sigma} \mathrm{Tr} \left[ \Pi \partial_\mu \Pi \partial_\nu \Pi \partial_\rho \Pi \partial_\sigma \Pi \right] \\
  & \supset \frac{2k}{15\pi^2 f_\pi^5} T_{[abcde]} \epsilon^{\mu\nu\rho \sigma} \pi^a \partial_\mu \pi^b \partial_\nu \pi^c \partial_\rho \pi^d \partial_\sigma \pi^e
  \,.
  \label{eq:WZWLagrangian}
}
Here, $k = N_C$ for $SU(N_C)$ and $2k = N_C$ for $SO(N_C)$ and $USp(N_C)$~\cite{Kamada:2017tsq}.
$T_{[abcde]}$ denotes the anti-symmetrization of five broken generators:
\eqs{
  T_{[abcde]} = \frac{1}{5!} \sum \mathrm{Tr} (T^{[a} T^b T^c T^d T^{e]}) \,.
  \label{eq:Tfive}
}
The four-point interactions arise at the order of $m_\pi^2/f_\pi^2$, while the five-point interaction is at the order of $m_\pi^5/f_\pi^5$. 
In accordance with the standard order-counting of the $\chi$PT using $p^2$ expansion in the chiral limit, the former is the LO contribution and the latter is the next-to-leading order (NLO) contribution in the non-chiral limit. 

A large coupling tends to be required for correct relic abundance in the SIMP scenarios.
In the dark-pion realization, the number-changing process arises from the WZW term, which is the higher order of the chiral expansion, with the velocity suppression.
Meanwhile, the $\chi$PT breaks down $m_\pi \simeq \Lambda_{\chi\mathrm{SB}}$, and the cutoff scale is expected to be
\eqs{
  \Lambda_{\chi\mathrm{SB}} \simeq \mathrm{min} \left( \frac{4 \pi f_\pi}{\sqrt{N_c}},\frac{4 \pi f_\pi}{\sqrt{N_f}} \right) \,.
  \label{eq:cutoffscale}
}
This is known as the scale estimated by na\"ive dimensional analysis (NDA)~\cite{Manohar:1983md,Georgi:1992dw} with taking into account the large-$N_c$ scaling and the large-$N_f$ scaling.
The large-$N_c$ scaling of $f_\pi$ and $\Lambda_{\chi\mathrm{SB}}$ is known to be $f_\pi^2 \simeq \mathcal{O}(N_c)$ and $\Lambda_{\chi\mathrm{SB}} \simeq \mathcal{O}(1)$~\cite{tHooft:1973alw,tHooft:1974pnl,Witten:1979kh,Witten:1979vv,Coleman:1980mx,Witten:1980sp}.
It is also known that some of the partial-wave amplitudes for the scattering process of $\pi\pi \to \pi\pi$ is proportional to $N_f$ in the large $N_f$ limit~\cite{Chivukula:1992nw,Chivukula:1992gi}.~\footnote{
  This discussion is based on the chiral symmetry breaking of $SU(N_f) \times SU(N_f) \to SU(N_f)$ in the original literature. 
  We find the similar $N_f$ dependence of the amplitude with the isospin singlet even for $SU(N_f)/SO(N_f)$ and $SU(N_f)/USp(N_f)$, and hence we expect that we get the same cutoff scale up to constant of order unity.
}
Since the pion self-coupling is required to be close to its na\"ive perturbative bound $m_\pi/f_\pi \lesssim \Lambda_{\chi\mathrm{SB}}/f_\pi$ in the SIMP scenario, the higher-order contributions of the $\chi$PT may affect the predictions of the relic abundance and the self-scattering cross section. 
Ref.~\cite{Hansen:2015yaa} has discussed the impacts of the higher-order contributions of the $\chi$PT in the context of the SIMP: in particular, the NLO and the next-to-next-to-leading order (NNLO) contributions.

%%%%%%%%%%%%%%%%%%%%%%%%%%%%%%%%%%%%%%%%%%%%%%%%%%%%%%%%%%%%%%%%%%%%%%%%%%%%%%%%%%%%%%%%
\section{Perturbative Unitarity \label{sec:unitarity}}

We discuss the perturbative unitarity of $\pi \pi \to \pi \pi$ and $\pi \pi \pi \to \pi \pi$ scattering processes.
The partial-wave decomposition of the invariant amplitude for elastic scattering process is given by%
\footnote{
  We use $32 \pi$ as a normalization instead of $16 \pi$ due to the identical particles.
}
\eqs{
  \mathcal{M}^{ab;cd}_{2 \to 2} & = 32 \pi \sum_R \sum_{\ell} (2 \ell+1) P_\ell(\cos\theta) T^R_\ell (s) P_R^{ab;cd} \,.
  \label{eq:pwdecomp}
}
Here, $P_\ell$ is the Legendre polynomial with $P_\ell(1) = 1$, $s$ denotes the collision energy, and $\theta$ is the scattering angle of the final state pions with respect to the collision axis. 
$P_R^{ab;cd}$ denotes the projection operators: 
the product of two pions is projected into the irreducible representation $R$ of residual global symmetry $H$.
$T^R_\ell (s)$ is the partial-wave elastic amplitude for the channel of a representation $R$.
The projection operators satisfy 
\eqs{
  \sum_{c',d'} P_{R}^{ab;c'd'} P_{R'}^{c'd';cd} & = \delta_{R R'} P_{R}^{ab;cd} \,, \quad
  (P_{R}^{ab;cd})^\ast = P_{R}^{cd;ab} \,, \quad
  \sum_{a,b} P_{R}^{ab;ab} = d_R \,.
}
Here, $d_R$ denotes the dimension of the representation $R$.
With this decomposition, the total cross section for $\pi^a \pi^b \to \pi^c \pi^d$ takes the form 
\eqs{
  \sigma^{ab;cd}_{2 \to 2} = \frac{32 \pi}{s} \sum_R \sum_{\ell} (2 \ell+1) |T^R_\ell (s)|^2 P_{R}^{ab;cd} \,.
}

The unitarity of the $S$-matrix imposes the optical theorem that relates the imaginary part of the invariant amplitude $\mathcal{M}_\mathrm{2 \to 2}$ for the forward scattering to the total cross section $\sigma_\mathrm{total}$.
Below the four-pion threshold, the optical theorem is given by
\eqs{
  \mathrm{Im} \mathcal{M}_{2 \to 2}
  = 2 \sqrt{s} p (\sigma_{2 \to 2} + \sigma_{2 \to 3}) \,.
}
Here, $p = \sqrt{s} \beta/2$ is the momentum of incoming particles with $\beta = (1-4m_\pi^2/s)^{1/2}$ denoting the velocity of the particles in the center of mass frame.
We use the relativistic formula for the two-body system ($p$ and $\beta$), while taking the non-relativistic limit of the three-body system. This is because the two-body system is semi-relativistic for the $2 \to 3$ process.
We include the inelastic cross section $\sigma_{2 \to 3}$ in the right-hand side since it is also sizable in the SIMP models, and it vanishes below the inelastic threshold.

\begin{figure}
	\centering
	\includegraphics[width=0.5\linewidth]{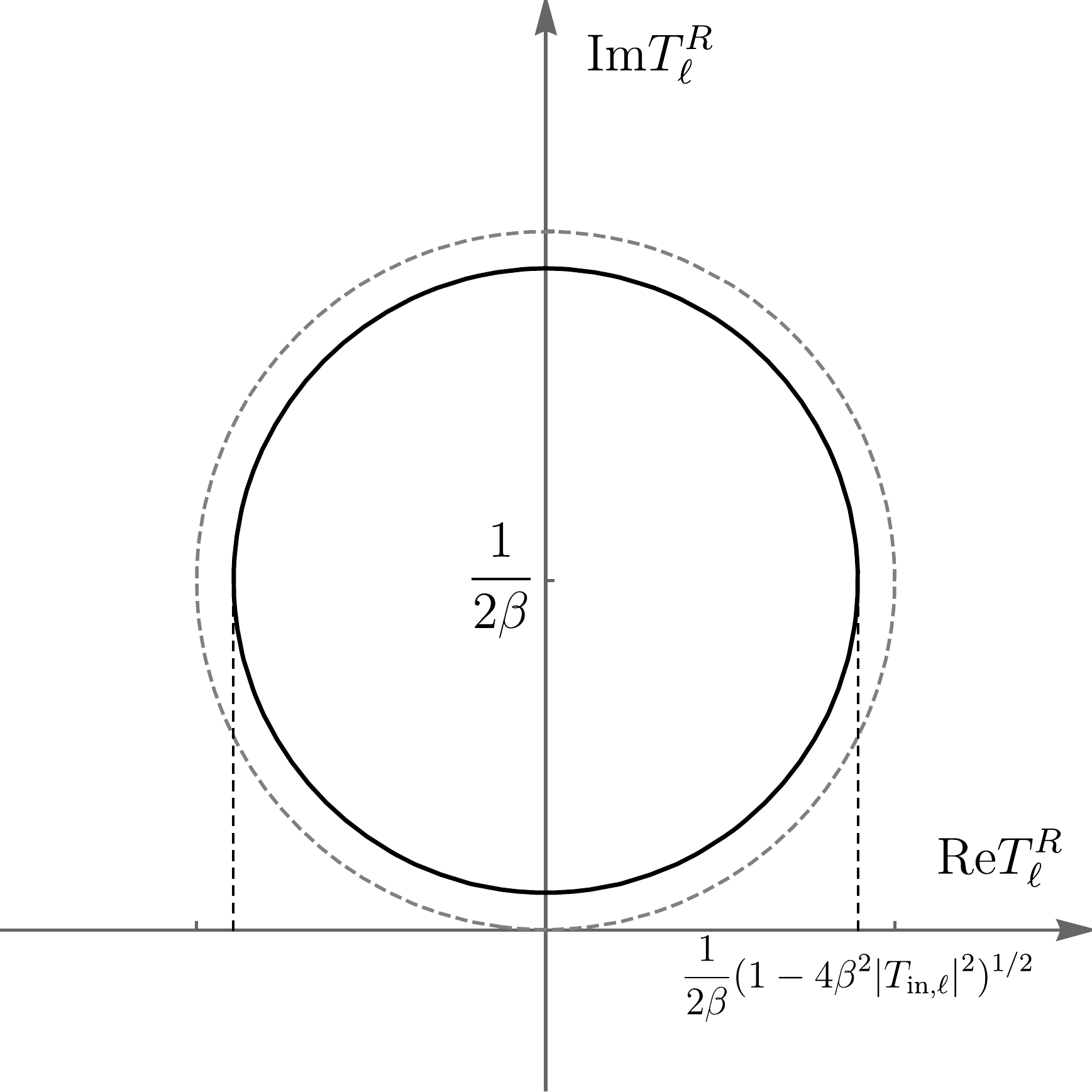}
	\caption{
  Unitarity circle on a complex plane of the partial-wave amplitude for the elastic scattering with a fixed partial-wave amplitude squared for the inelastic scattering.
  Unitarity circle without the inelastic scattering is shown as the gray-dashed line.
	}
	\label{fig:Ucircle1}
\end{figure}
 
We define the partial-wave amplitude for the $\pi \pi \to \pi \pi \pi$ process in the non-relativistic limit of the three-body final state.
It is challenging to give the explicit form of the decomposition of the three-body final state into its irreducible representations, in other words to construct the explicit form of the projection operator $P_R^{abc;de}$.
Meanwhile, the cross section can be decomposed into the irreducible representations of the initial state.
The cross section for the $\pi \pi \to \pi \pi \pi$ process may be written in a similar way to the $\pi \pi \to \pi \pi$ process. 
The elastic cross section and the inelastic cross section averaged over all initial states are given by
\eqs{
  \sigma_{2 \to 2} & 
  \equiv \frac{1}{N_\pi^2} \sum_{a,b} \sigma^{ab;ab}_{2 \to 2} 
  = \frac{32 \pi}{s} \frac{1}{N_\pi^2} \sum_{R}\sum_{\ell} (2 \ell+1) |T^R_\ell|^2 d_R \,, \\
  \sigma_{2 \to 3} & = \frac{32 \pi}{s} \frac{1}{N_\pi^2} \sum_{R}\sum_{\ell} (2 \ell+1) |T^R_{\mathrm{in}, \ell}|^2 d_R \,.
  \label{eq:2to2_2to3}
}
Here, we define the square of the inelastic amplitude $|T^R_{\mathrm{in}, \ell}|^2$.
When we include the $\pi \pi \to \pi \pi \pi$ inelastic scattering process, the optical theorem for each partial-wave amplitude takes the form,
\eqs{
  \mathrm{Im}(T^R_\ell) = \beta (|T^R_\ell|^2 + |T^R_{\mathrm{in}, \ell}|^2) \,.
  \label{eq:opticaltheorem}
}

Let us apply this condition to perturbative analysis in the case that inelastic scattering is negligible.
The optical theorem relates a fixed-order amplitude to the imaginary part of the higher-order amplitude: the tree-level partial-wave amplitude for the elastic scattering $T_\ell^\mathrm{tree}$ is real, and is related to the imaginary part of the one-loop amplitude.
We give the perturbative analysis of $\lambda \phi^4$ theory in \cref{app:phi4}, and discuss this point explicitly.
In other words, the fixed-order amplitude does not satisfy \cref{eq:opticaltheorem}.
Hence, it is important to unitarize the amplitude in order to satisfy the optical theorem automatically as the pion self-coupling is large.

The optical theorem places upper bounds on the partial-wave amplitudes.
\cref{fig:Ucircle1} shows the unitarity circle on a complex plane of the partial-wave amplitude for the elastic scattering.
The gray dashed line depict the unitarity circle without the inelastic scattering; its center is located at $(0,1/2\beta)$ and its radius is $1/2\beta$. 
The radius shrinks in the presence of the inelastic scattering as depicted as the solid-line circle, and the single point $(0,1/2\beta)$ is allowed once the inelastic scattering saturates the unitarity bound ($|T^R_{\mathrm{in}, \ell}|^2 = 1/4 \beta^2$).
The upper bound on the partial-wave amplitude for the elastic scattering is given by $|T^R_\ell| \leq 1/\beta$ as shown in the figure.
The equality holds when $\mathrm{Re} T^R_\ell = 0$, $|T^R_{\mathrm{in}, \ell}|^2 = 0$, and $\mathrm{Im} T^R_\ell= 1/\beta$, corresponding to the top of the gray-dashed circle.
The perturbative unitarity is often imposed as $|\mathrm{Re} T^R_\ell| \leq 1/2\beta$. 
The equality holds when the imaginary part satisfies $\mathrm{Im} T^R_\ell = 1/2\beta$ and the inelastic scattering is negligible.

We discuss the unitarity bound on the pion self-coupling based on the partial-wave unitarity of tree-level $\pi^a(p_1) \pi^b(p_2) \to \pi^c(p_3) \pi^d(p_4)$ self-scattering cross section.
At the LO terms of $\chi$PT, the four-point interaction is given by \cref{eq:4pi}, and the tree-level amplitude of the process is obtained as follows:
\eqs{
  \mathcal{M}^{ab;cd}_{2 \to 2} (p_1, p_2; p_3, p_4) & = 
  \frac{m_\pi^2}{f_\pi^2} c_{abcd} - \frac{s}{f_\pi^2} r_{abcd} - \frac{t}{f_\pi^2} r_{acbd} - \frac{u}{f_\pi^2} r_{adbc} \,.
  \label{eq:selfscatt_amp}
}
Here, the Mandelstam variables are defined by $s = (p_1+p_2)^2 \,, t = (p_1 - p_3)^2$\,, and $u = (p_1 - p_4)^2$.
The singlet channel for $\pi \pi \to \pi \pi$ has the larger coefficient compared to the others in an $SU(N_f) \times SU(N_f)/SU(N_f)$ with a large $N_f$ limit, and the $s$-wave amplitude for this channel takes the form,
\eqs{
  T_{\ell = 0}^{R=1} = \frac{N_f}{8\pi} \frac{m_\pi^2}{f_\pi^2} \,,
}
where we take the non-relativistic limit of two-body system, $s = 4 m_\pi^2 + 4 m_\pi^2 \beta^2$ with small $\beta$.
Once the pion self-coupling is in the range of $m_\pi/f_\pi \gtrsim \sqrt{4\pi/N_f \beta}$, the partial-wave amplitude goes beyond the unitarity bound.
The elastic scattering cross section can exceed the perturbative unitarity bound when the pion self-coupling within the following range:
\eqs{
  \sqrt{\frac{4\pi}{\beta N_f}} \lesssim \frac{m_\pi}{f_\pi} \lesssim \mathrm{min} \left( \frac{4 \pi}{\sqrt{N_c}},\frac{4 \pi}{\sqrt{N_f}} \right)\,. 
  \label{eq:couplingrange}
}
This upper bound originates from the condition where tree and one-loop contributions are compatible to each other. 

%%%%%
We also discuss the unitarity bound on the partial-wave amplitude for the inelastic scattering, $\pi^a(p_1) \pi^b(p_2) \to \pi^c(p_3) \pi^d(p_4) \pi^e (p_5)$.
The WZW term (\ref{eq:WZWLagrangian}) provides the five-pion interaction that induces the inelastic scattering.
The invariant amplitude at the tree-level is given by 
\eqs{
  \mathcal{M}_\mathrm{tree}^{abcde}(p_1,p_2,p_3,p_4,p_5) = \frac{16 k}{\pi^2 f_\pi^5} T_{[abcde]} \epsilon^{\mu\nu\rho\sigma} p_{2\mu} p_{3\nu} p_{4\rho} p_{5\sigma} \,.
  \label{eq:5ptvertex}
}
Here, $T_{[abcde]}$ is defined in \cref{eq:Tfive}.
We compute the $\pi \pi \to \pi \pi \pi$ cross section in the non-relativistic limit of the three-body final state, and we extract the partial-wave amplitude for the square of the inelastic scattering by comparing the cross section with \cref{eq:2to2_2to3}.
Since the amplitude has only $p$-wave component at the tree level even for the relativistic limit, only antisymmetric representations in the irreducible decomposition of initial two-pion states contribute to the amplitude. 
We find that the square of the partial-wave amplitude takes the form, 
\eqs{
  (T^R_{\mathrm{in},\ell = 1})^2
  = \left( \frac{1}{4\sqrt3 \pi} \right)^7 4 k^2 \beta v^8 \left( \frac{m_\pi}{f_\pi} \right)^{10} \frac{t_R^2}{d_R} \,,
  \label{eq:pw_inelastic}
}
where we take the non-relativistic limit of the three-body system, $s = 9 m_\pi^2 + 6 m_\pi E$ with the relative energy $E = \bar \mu v^2/2$ and the three-body reduced mass $\bar \mu = m_\pi/\sqrt{3}$.
Meanwhile, $\beta^2 = 1 - 4 m_\pi^2/s \simeq 5/9 + 4 \sqrt3 v^2/81$ is semi-relativistic.
$t_R^2$ denotes a group theoretical factor, which scales as $N_f^5$.
$2 \beta$ and $v$ correspond to the relative velocities of the two-body initial state and of the three-body final state, respectively. 
In general, the inelastic partial-wave amplitude squared has the velocity dependence of $(T^R_{\mathrm{in},\ell})^2 \propto \beta^{2 \ell-1} v^{4 + 2 \lambda}$ with $\lambda$ denoting hyperangular quantum number, and the $2 \to 3$ cross section may scale as $v^6$ ($\lambda = 1$) at the leading order of the $p$-wave cross section.
In the dark-pion realization, the partial-wave amplitude is the next-leading order of the $p$-wave amplitude ($\lambda = 2$).

The unitarity bound on the partial-wave amplitude, $1/2\beta$, places an upper bound on the pion self-coupling $m_\pi/f_\pi$. 
The smallest representation in dimension provides the largest partial-wave amplitude.
In an $SU(N_f) \times SU(N_f)/SU(N_f)$ with a large $N_f$ limit, $d_R$ for the adjoint representation scales as $N_f^2$, while $d_R$ for the larger representations scale as $N_f^4$.
The partial-wave amplitude for the adjoint representation gives stronger bound on $m_\pi/f_\pi$ in a large $N_f$ limit, and it takes the form,
\eqs{
  \left( \frac{m_\pi}{f_\pi} \right)^{10} \lesssim
  \frac{3(8\sqrt3 \pi)^7}{4} \frac{1}{N_f^3 N_c^2} \frac{1}{\beta^3 v^8} \,.
}
This bound on $m_\pi/f_\pi$ is numerically weaker than the perturbative unitarity bound obtained from the $2 \to 2$ scattering for the velocity $\beta \simeq v \simeq 1$.

%%%%%

Next, we discuss the partial-wave unitarity of the thermally-averaged cross section for the $\pi\pi\pi \to \pi\pi$ process.
We will give detailed computations in \cref{app:crosssection}. 
When the invariant amplitude for the $\pi\pi \to \pi\pi\pi$ has time-reversal invariance%
\footnote{
  This is ensured by the $CPT$ invariance in the dark pion case. 
}
, the cross section for $\pi\pi\pi \to \pi\pi$ in the non-relativistic limit of the three-body system ($s = 9 m_\pi^2 +\sqrt3 m_\pi^2 v^2$) is also written by the use of the partial-wave amplitude squared $|T^R_{\mathrm{in},\ell}|^2$, and hence the cross section takes the form:
\eqs{
  \sigma_{3 \to 2} v^2 (\bar k)
  = \frac{192 \sqrt{3} \pi^2}{\bar k^4 m_\pi} \frac{1}{N_\pi^3} \sum_{R}\sum_{\ell} (2 \ell+1) 4 \beta^2 |T^R_{\mathrm{in},\ell} (9m_\pi^2)|^2 d_R \,,
  \label{eq:crosssection_3to2}
}
with the Jacobi momentum $\bar k \equiv \overline \mu v$. 
The partial-wave unitarity $|T^R_{\mathrm{in},\ell = 1}|^2 \leq 1/4\beta^2$ places an upper bound on the cross section. 
\eqs{
  \sigma_{3 \to 2:\mathrm{uni} } v^2 (\bar k)
  = 3 \times \frac{192 \sqrt{3} \pi^2}{\bar k^4 m_\pi} \frac{1}{N_\pi^3} \sum_{R} d_R \,.
}
Here, a factor of 3 arises from the fact that $\pi\pi\pi \to \pi\pi$  scattering involves only $p$-wave.
The thermally-averaged cross section is obtained by integrating the cross section over $\overline k$ with the Maxwell-Boltzmann distribution.
\eqs{
  \langle \sigma_{3 \to 2} v^2  \rangle_\mathrm{uni} 
  = 3 \times \frac{12 \sqrt3 \pi^2 x^2}{m_\pi^5} \frac{1}{N_\pi^3} \sum_{R} d_R\,.
  \label{eq:unitarity_3to2_therm}
}
Here, $x = m_\pi/T$ with $T$ denoting temperature.
This bound differs by a factor of $1/4$ from what is obtained in Refs.~\cite{Kuflik:2017iqs,Namjoo:2018oyn}.
We use the upper bound on the partial-wave amplitude for the inelastic scattering given by $1/2\beta$ in the presence of the elastic scattering, while the upper bound for the elastic scattering ($1/\beta$) has been used for constraining the thermally-averaged cross section in Refs.~\cite{Kuflik:2017iqs,Namjoo:2018oyn}.%
\footnote{
  We obtain the upper bounds by the use of an independent way from Refs.~\cite{Kuflik:2017iqs,Namjoo:2018oyn,Bhatia:2020itt}.
  As shown in \cref{app:crosssection}, we directly compute the upper bound on the hyperangle averaged cross section in the non-relativistic limit of the $N$-body initial states, and then we take the thermal average of it. 
  Meanwhile, the authors in Refs.~\cite{Kuflik:2017iqs,Namjoo:2018oyn,Bhatia:2020itt} obtain the upper bound on the thermally-averaged $n \to 2$ cross section by using the detailed balance condition. 
  Our result is consistent with Refs.~\cite{Kuflik:2017iqs,Namjoo:2018oyn} up to a factor of $1/4$ due to maximizing the partial-wave amplitude in the presence of the elastic amplitude, but our result disagrees with the results in Ref~\cite{Bhatia:2020itt}.
}
The summation runs only over the antisymmetric representations since the square of the inelastic amplitude is nonzero only for the antisymmetric representations.

Let us now consider the $\pi\pi\pi \to \pi\pi$ scattering at the leading contribution in the chiral Lagrangian. 
We have given the partial-wave amplitude for the $\pi \pi \to \pi \pi \pi$ cross section at the leading contribution for the process in \cref{eq:pw_inelastic}.
The thermally-averaged cross section for the $\pi\pi\pi \to \pi\pi$ process is%
\footnote{
  Our result differs from the original literature~\cite{Hochberg:2014kqa} by a factor of $1/3$. 
  The difference arises from mismatching frames of reference; the reference frame for computing the cross section $\sigma_{3 \to 2}$ is generally different from the frame for the thermal averaging.
  In particular, the cross section explicitly depends on momenta of particles in the dark pion SIMP. 
  As for the thermal averaging, the cross section should be written in the Galilean invariant way.
  We discuss this point in \cref{app:crosssection} in detail. 
}
\eqs{
  \langle \sigma_{3 \to 2} v^2 \rangle = \frac{5 \sqrt5 k^2 m_\pi^5}{6 \pi^5 x^2 f_\pi^{10} } \frac{t^2}{N_\pi^3} \,.
}
where $t^2 \equiv \sum (T_{[abcde]})^2$.
We show the explicit forms of $t^2$ in different symmetry structures in \cref{app:group}.
$t^2/N_\pi^3$ approaches to $1/N_f$ at the large-$N_f$ limit.

The thermally-averaged cross section will exceed the unitarity bound given by \cref{eq:unitarity_3to2_therm} in the large $m_\pi/f_\pi$ limit. 
Once we take the na\"ive perturbative bound of the pion self-coupling, $m_\pi/f_\pi \simeq 4 \pi$, the cross section takes the form
\eqs{
  \langle \sigma_{3 \to 2} v^2 \rangle \simeq \frac{5 \sqrt5 (4 \pi)^5}{6 x^2 m_\pi^5} \frac{4}{3 N_f}\,,
}
in an $SU(N_f) \times SU(N_f)/SU(N_f)$ with a large $N_f$ limit.
The annihilation cross section is safe from the unitarity violation at large $x$, namely at low temperature, since the cross section is suppressed $x^{-2}$ while the unitarity bound is proportional to $x^{2}$.
The cross section, however, will be beyond the unitarity bound at small $x$ even for the pion self-coupling below the na\"ive perturbative bound.

%%%%%%%%%%%%%%%%%%%%%%%%%%%%%%%%%%%%%%%%%%%%%%%%%%%%%%%%%%%%%%%%%%%%%%%%%%%%%%%%%%%%%%%%
\section{Improved Cross Section \label{sec:improvement}}

At the large pion self-coupling, both of $\pi \pi \to \pi \pi$ and $\pi \pi \pi \to \pi \pi$ scattering processes at tree-level can be beyond the perturbative unitarity even for the coupling within the na\"ive perturbative range.
In this section, we propose the improvement of the partial-wave amplitude to take into account the partial-wave unitarity of the cross section. 
We may use perturbative analysis for the pion self-coupling within the na\"ive perturbative range given by \cref{eq:couplingrange}, but we have to modify the amplitude to satisfy the optical theorem.
In particular, we need to incorporate the imaginary part of the amplitude at the LO into an improved amplitude. 
We define the improved amplitude whose real and imaginary parts appropriately reproduce the original amplitude at the LO as follows: 
\eqs{
  T^{\mathrm{imp},R}_\ell
  &= \frac{T^{\mathrm{tree},R}_\ell + i \beta (T^{\mathrm{tree},R}_{\mathrm{in}, \ell})^2}{1 - i \beta [T^{\mathrm{tree},R}_\ell + i \beta (T^{\mathrm{tree},R}_{\mathrm{in}, \ell})^2]} \,,
  \\
  |T^{\mathrm{imp},R}_{\mathrm{in}, \ell}|^2
  &= \frac{(T^{\mathrm{tree},R}_{\mathrm{in}, \ell})^2}{[1+(\beta T^{\mathrm{tree},R}_{\mathrm{in}, \ell})^2]^2 + (\beta T^{\mathrm{tree},R}_\ell)^2} \,. 
   \label{eq:improved_amplitude}
}
This improved partial-wave amplitude automatically satisfies \cref{eq:opticaltheorem} at any $m_\pi/f_\pi$: namely, $\mathrm{Im}(T^\mathrm{imp}) = \beta (|T^\mathrm{imp}|^2 + |T^\mathrm{imp}_\mathrm{in}|^2)$ for each $R$ and $\ell$.
In the absence of the inelastic scattering amplitude squared $(T^\mathrm{tree}_\mathrm{in})^2$, this improvement of the partial-wave amplitudes coincides with the ``self-healing'' mechanism in which the perturbative unitarity violation is cured by resumming multiple rescattering processes~\cite{Aydemir:2012nz}.

\begin{figure}
	\centering
	\includegraphics[width=\linewidth]{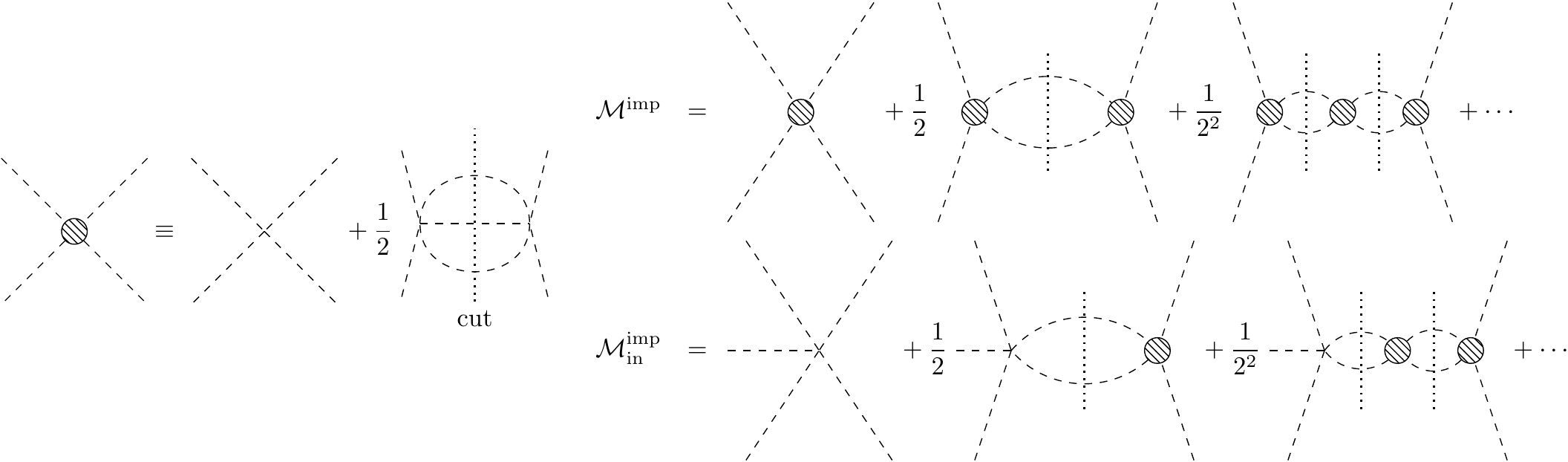}
	\caption{
  Diagrammatic interpretation of our improvement in $2 \to 2$ and $2 \to 3$ amplitudes.
  (\textit{Left}): an effective vertex (denoted by a meshed blob) includes the tree-level four-point vertex and the one-loop vertex induced by the tree-level five-point vertex.
  (\textit{Right}): the improved amplitude for $2 \to 2$ and $2 \to 3$ processes.
	}
	\label{fig:resummation}
\end{figure}

Our improvement amplitudes are not either unique or complete to remedy the situation violating the unitarity condition, but are associated with resumming the multiple rescattering processes that arise not only from the four-point vertex but also from the tree-level five-point vertex.
\cref{fig:resummation} shows diagrammatic interpretation of invariant amplitudes with our improved partial-wave amplitudes.
A blob in this figure denotes the effective four-point vertex including the tree-level four-point vertex and the one-loop contributions with cuts and the insertion of the tree-level five-point vertex.
The vertical dotted-lines depict the cutting of intermediate propagators with replacement:
\eqs{
  \frac{i}{k^2 - m_\pi^2 + i \varepsilon}
  \to - 2 \pi i \delta (k^2 - m_\pi^2) \,, 
}
where $k$ denotes the loop momentum.
These improved amplitudes correspond to a geometric sum of multiple self-rescattering processes, in other words a geometric sum of diagrams where all loops are cut as shown in \cref{fig:resummation}.
This improved amplitude consists only of the tree-level amplitude and cut diagrams, and hence this procedure is not appropriate when full one-loop corrections are important.

\subsection{Analytic Properties}

Before showing the impact of the improved amplitude on the SIMP phenomenology, we discuss the analytic properties of the improved amplitudes, such as the scattering phase shift and the poles. 
The phase shift $e^{2 i \delta_\ell^R} \equiv 1 + 2 i \beta T^R_\ell$ may be written as
\eqs{
  \cot \delta_\ell^R
  = \frac{1 + i \beta T^R_\ell}{\beta T^R_\ell} \,.
}
Meanwhile, the phase shift in the low-energy limit can be expanded 
\eqs{
  p^{2 \ell + 1} \cot \delta_\ell^R \simeq - \frac{1}{a_\ell^{2 \ell+1}} + \frac{p^2}{2r_{e,\ell}^{2 \ell-1}} \,,
}
where $p = \sqrt{s} \beta/2$ denotes the relative momentum of the system. 
$a_\ell$ and $r_{e, \ell}$ are known as the scattering length and the effective range, respectively.
The phase shift is generically complex in the presence of the inelastic scattering, such as DM annihilation \cite{Braaten:2013tza,Chu:2019awd}.

Using our improved partial-wave amplitude, the phase shift is given by the tree-level partial-wave amplitude as follows.
\eqs{
  \cot \delta_\ell^{\mathrm{imp},R}
  = \frac{1}{\beta T^{\mathrm{tree},R}_\ell + i \beta^2 (T^{\mathrm{tree},R}_{\ell, \mathrm{in}})^2} \,.
}
The phase shift with our improved amplitude is indeed complex since the partial-wave amplitudes are real at tree level. 
The imaginary part of the phase shift leads to shrinking the radius of the unitarity circle of elastic scattering as shown in \cref{fig:Ucircle1}.
As for $s$-wave amplitudes, the elastic amplitude is proportional to $\beta^0$ in the non-relativistic limit and there is no inelastic scattering. 
On the other hand, the $p$-wave elastic amplitude is proportional to $\beta^2$. 
Therefore, we confirm that $\beta^{2 \ell + 1} \cot \delta_\ell^{\mathrm{imp},R}$ is analytic at $\beta = 0$ even in our improvement procedure for the application to the dark pion SIMP models.
The square of the inelastic amplitude vanishes $\beta \lesssim \sqrt{5}/3$ due to the kinematics, while the $\beta$-dependence of the square of the inelastic amplitude is compatible with the expression of the effective range theory. 
We note that, assuming that the imaginary part of the phase shift is negligible compared to the real part, the phase shift is given by the tree-level partial-wave amplitude as $\mathrm{Re} (\cot \delta_\ell^{\mathrm{imp}, R}) \simeq (\beta T_\ell^{\mathrm{tree},R})^{-1}$ and $\mathrm{Im} (\cot \delta_\ell^{\mathrm{imp}, R}) \simeq - (T^{\mathrm{tree},R}_{\ell, \mathrm{in}}/ T_\ell^{\mathrm{tree},R})^{2}$.
Under the Born approximation, the phase shift based on the tree-level amplitude is given by $\delta_\ell \simeq \beta T^\mathrm{tree}_\ell$: i.e., $\cot \delta_\ell \simeq (\beta T^\mathrm{tree}_\ell)^{-1}$, which is reproduced by the expression of $\delta_\ell^{\mathrm{imp},R}$. 

Our improved amplitudes would involve a pole.
There is no inelastic channel for $s$-wave amplitudes, and the amplitude is proportional to $\beta^0$ at the LO. 
Therefore, the pole would appear at 
\eqs{
  \beta_\mathrm{pole} = - \frac{i}{T^{\mathrm{tree},R}_{\ell=0}} \,.
} 
It is indicated that $|T^{\mathrm{tree},R}_{\ell=0}| \gtrsim 1$, i.e., $|\beta| \lesssim 1$, in the non-relativistic limit.
The bound state ($\mathrm{Im} \beta_\mathrm{pole} > 0$) or the virtual level ($\mathrm{Im} \beta_\mathrm{pole} < 0$) would appear depending on the sign of the partial-wave amplitudes.

As for the $p$-wave amplitudes, the amplitude for the elastic scattering is proportional to $\beta^2$ at the LO: we set the $\beta$-independent part of the amplitude as $T^{\mathrm{tree},R}_{\ell=1} = \beta^2 t^{\mathrm{tree},R}_{\ell=1}$.
The partial-wave amplitude squared for inelastic scattering $(T^{\mathrm{tree},R}_\ell)^2$ is proportional to $\beta$, and hence we set the $\beta$-independent part as $(T^{\mathrm{tree},R}_\ell)^2 = \beta (t^{\mathrm{tree},R}_\ell)^2$.
As far as the inelastic channels is negligible, the pole would appear at 
\eqs{
  \beta_\mathrm{pole}^3 = - \frac{i}{t^{\mathrm{tree},R}_{\ell=1}} \,.
}
We can find a pole on the imaginary axis of $\beta$, and hence the pole corresponds to the bound state ($t^{\mathrm{tree},R}_{\ell=1} > 0$) or the virtual level ($t^{\mathrm{tree},R}_{\ell=1} < 0$).

\subsection{Application to dark pion SIMP}

Now, we discuss the impact of the improvement on the SIMP models. 
Using the tree-level amplitudes given by \cref{eq:4pi,eq:WZWLagrangian}, we obtain the improved amplitudes for $\pi \pi \to \pi \pi$ and $\pi \pi \pi \to \pi \pi$ processes.
The improved amplitudes for $2 \to 2$ scattering coincide with the tree-level amplitudes in the $\beta \to 0$ limit, when the inelastic channel closes.
Hence, the improvement has a low impact on the pion $2 \to 2$ self-scattering cross section in cosmic structures due to the DM velocity of $\sim 10^{-2}$ at maximum.
Meanwhile, this improvement procedure will be efficient when the pions are semi-relativistic. 
The tree-level partial-wave amplitudes scale as $T^\mathrm{tree} \propto (m_\pi/f_\pi)^2$ and $(T^\mathrm{tree}_{\mathrm{in}})^2 \propto (m_\pi/f_\pi)^{10}$, and hence the inelastic scattering more significantly affects the improvement at the large $m_\pi/f_\pi$.

\begin{figure}
	\centering
	\includegraphics[width=0.4\linewidth]{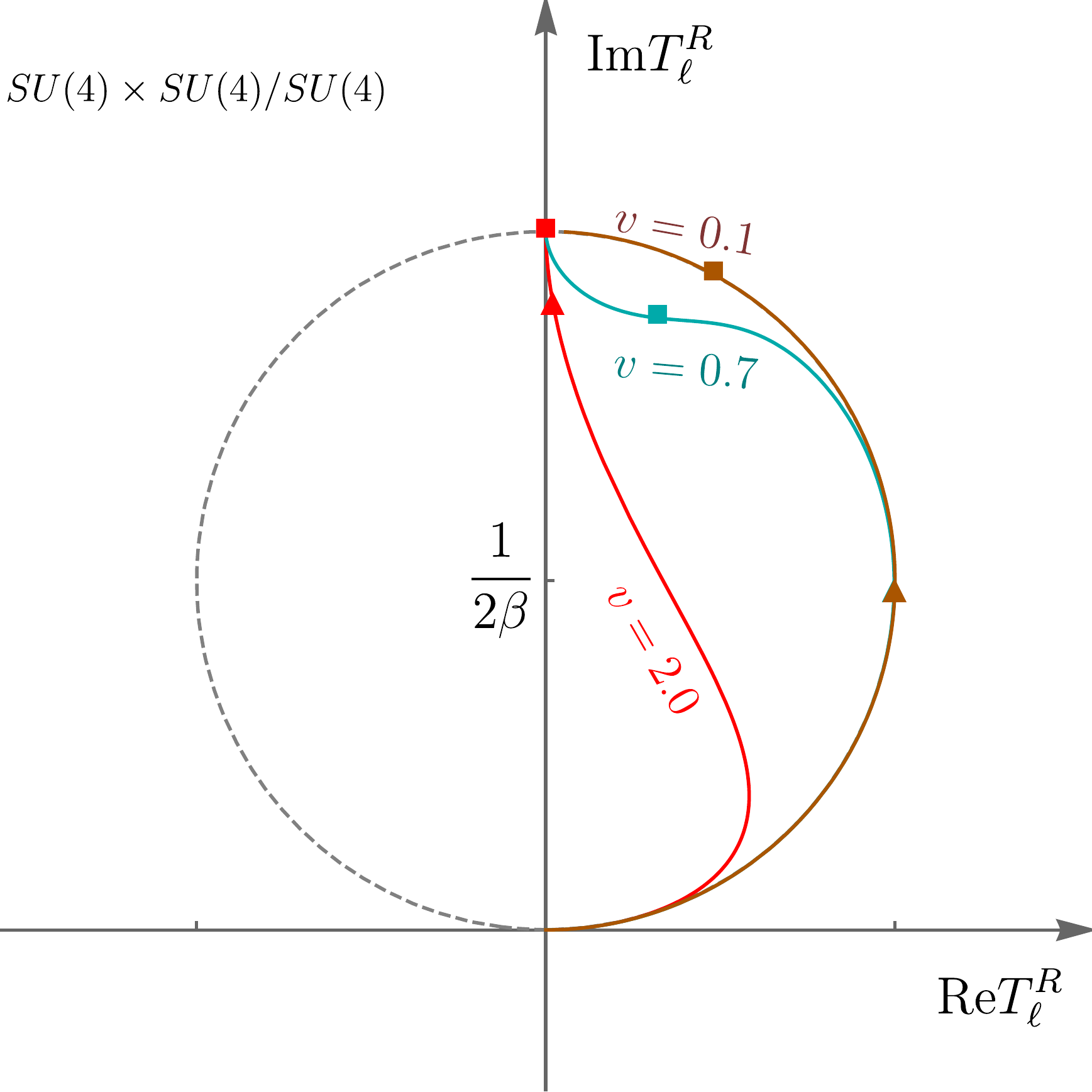}
	\caption{
  Trajectories of the improved partial-wave amplitude on the complex plane of the amplitude for the elastic scattering as $m_\pi/f_\pi$ increased.
  The gray-dashed line corresponds to the unitarity circle (without the inelastic scattering), whose center is at $(0,1/2\beta)$ and radius is $1/2\beta$.
  We assume $SU(N_f) \times SU(N_f)/SU(N_f)$ with $N_f = 4$ and $N_c = 4$, and the adjoint representation for $R$.
  Each line corresponds to fixed $v$: $v = 0.1$ (brown), $v = 0.7$ (cyan), and $v = 2.0$ (red).
  There are two markers on each line, which correspond to $m_\pi/f_\pi = 4 \pi/\sqrt{N_c}$ and $4 \pi$. 
	}
	\label{fig:Ucircle2}
\end{figure}

We plot the improved $p$-wave amplitude for the elastic scattering, $T^{\mathrm{imp}}$, on a complex plane in \cref{fig:Ucircle2} for the adjoint channel of $SU(4) \times SU(4)/ SU(4)$. 
Both elastic and inelastic scattering processes exist at the tree-level in this channel.
In the figure, we take different velocities $v$ of the three-body initial state: $v = 0.1$ (brown), $v = 0.7$ (cyan), and $v = 2.0$ (red).%
\footnote{
  One can estimate the maximum relative velocity of $n$-body identical-particle collision as $v \leq n^{n/2(n-1)}$ by assuming that each particle comes from respective vertex of a regular $n$-sided polygon with the same velocity. 
  In particular, the relative velocity can be $v \simeq 2.3$ in the three-body collision. 
}
Once we specify the symmetry structure and the channel, the improved amplitude is a function depending only on $m_\pi/f_\pi$ and velocity $v$.
We note that the relative velocity of the two-body final state is fixed as $\beta = \sqrt{5}/3$ in the non-relativistic limit of the three-body initial state.
As $m_\pi/f_\pi$ increases from zero to infinity, the amplitude for fixed $v$ moves from bottom to top in the figure.
We put marks on each lines at $m_\pi/f_\pi = 4 \pi/\sqrt{N_c}$ ($\blacktriangle $) and $4 \pi$ ($\blacksquare$). 
The improved amplitude $T^{\mathrm{imp}}$ is on the unitarity circle (with the radius of $1/2\beta$) when the inelastic scattering is negligible. 
For small $m_\pi/f_\pi$, therefore, each line follows the unitarity circle since the square of the tree-level partial-wave amplitude for the inelastic scattering is suppressed by a higher power of $m_\pi/f_\pi$ than the tree-level partial-wave amplitude for the elastic scattering.
$T^{\mathrm{imp}}_\mathrm{in}$ can be comparable to $T^{\mathrm{imp}}$ at a certain point of $m_\pi/f_\pi$, and thus the radius from the center of the circle shrinks as we see in \cref{fig:Ucircle1}. 
Since the square of the partial-wave amplitude for the inelastic scattering, $(T_{\mathrm{in,},\ell=1}^R)^2$, is proportional to $v^8$ as shown in \cref{eq:pw_inelastic}, the path goes inside the circle for smaller $m_\pi/f_\pi$ as the velocity $v$ gets larger.
The improved partial-wave amplitudes approach to $(0,1/\beta)$ for sufficiently large $m_\pi/f_\pi$. 
As shown in \cref{eq:improved_amplitude}, $T^{\mathrm{imp}} \to i/\beta$ and $T^{\mathrm{imp}}_\mathrm{in} \to 0$ when the tree-level inelastic scattering gets increased. 

\begin{figure}
	\centering
	\includegraphics[width=0.45\linewidth]{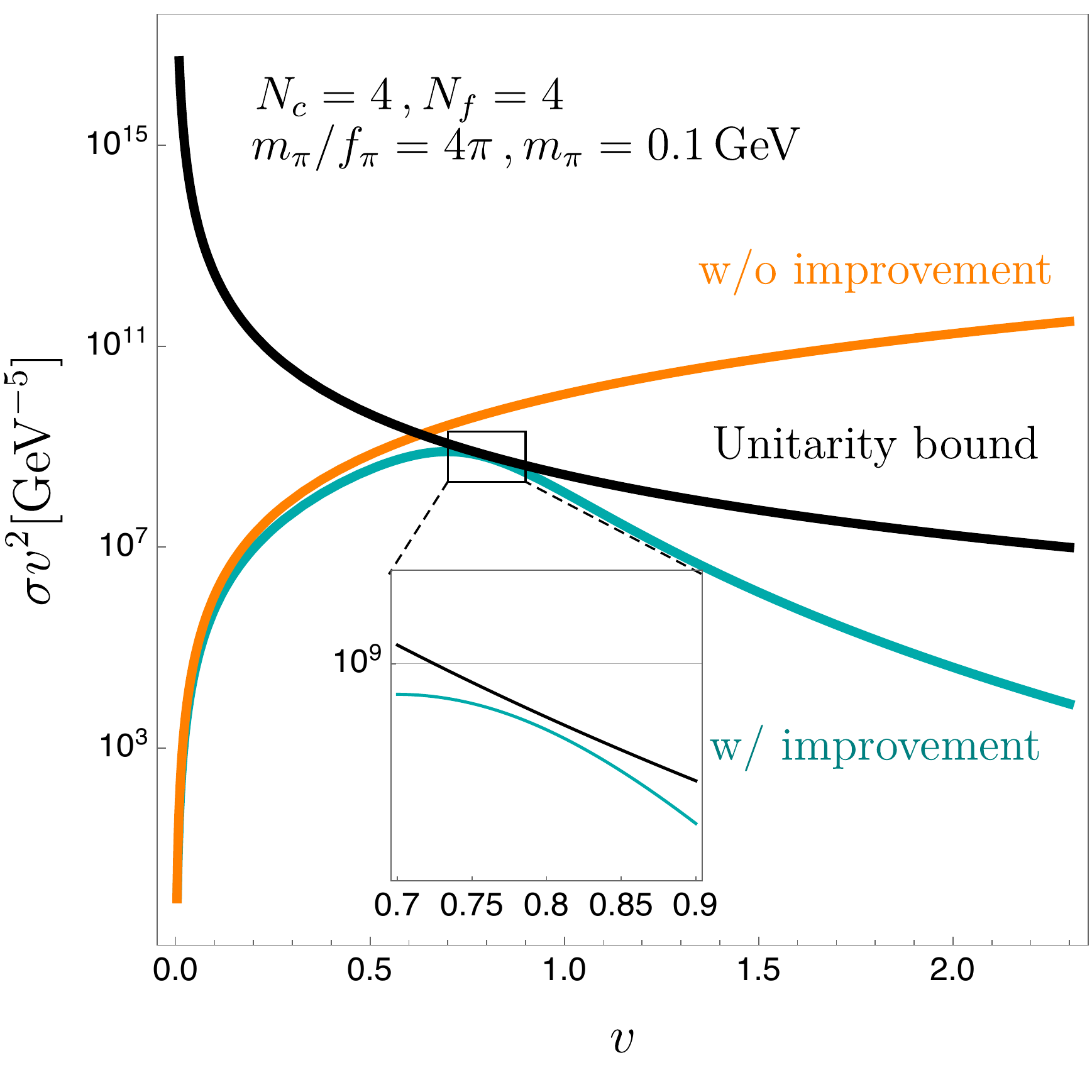}
	\includegraphics[width=0.45\linewidth]{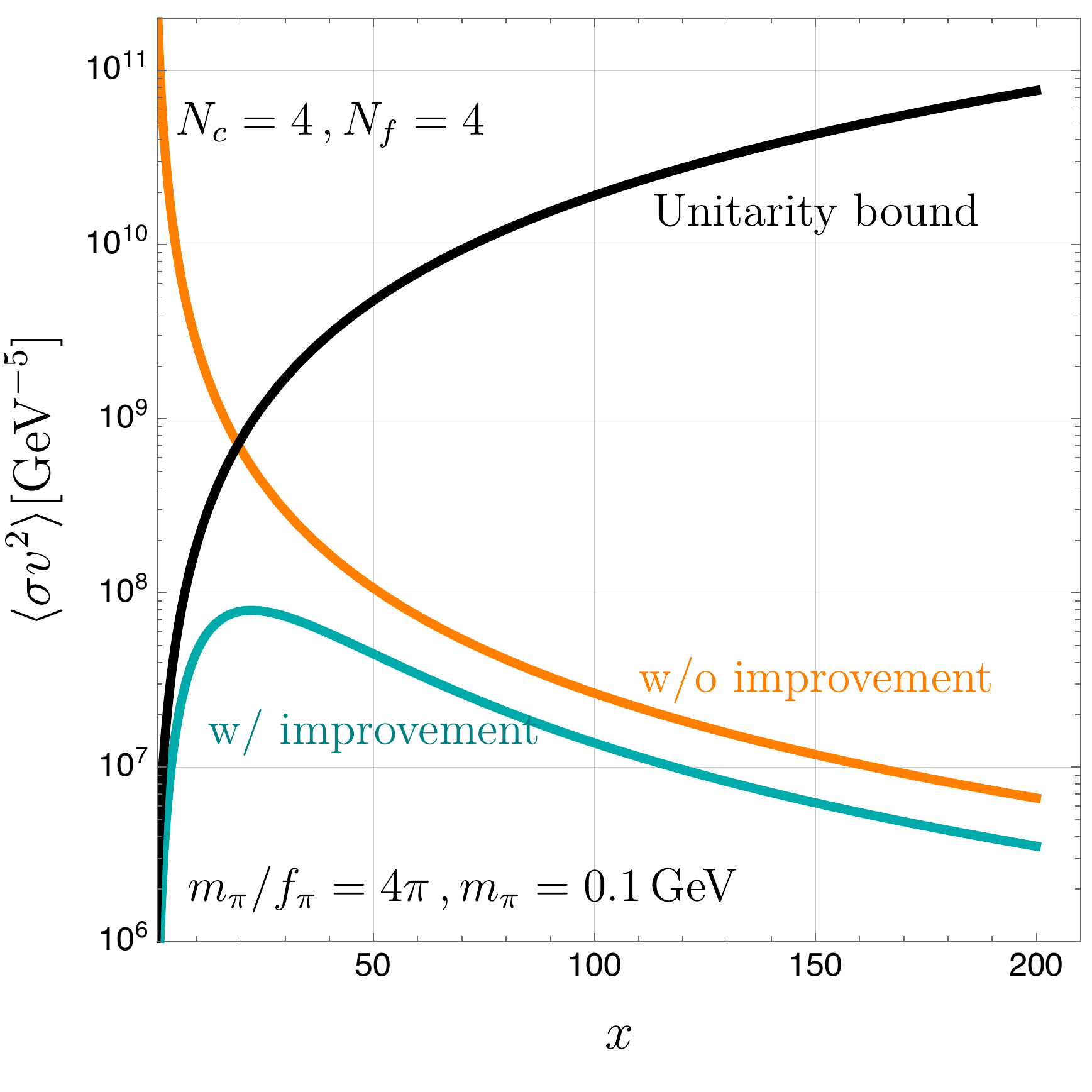}
  \includegraphics[width=0.45\linewidth]{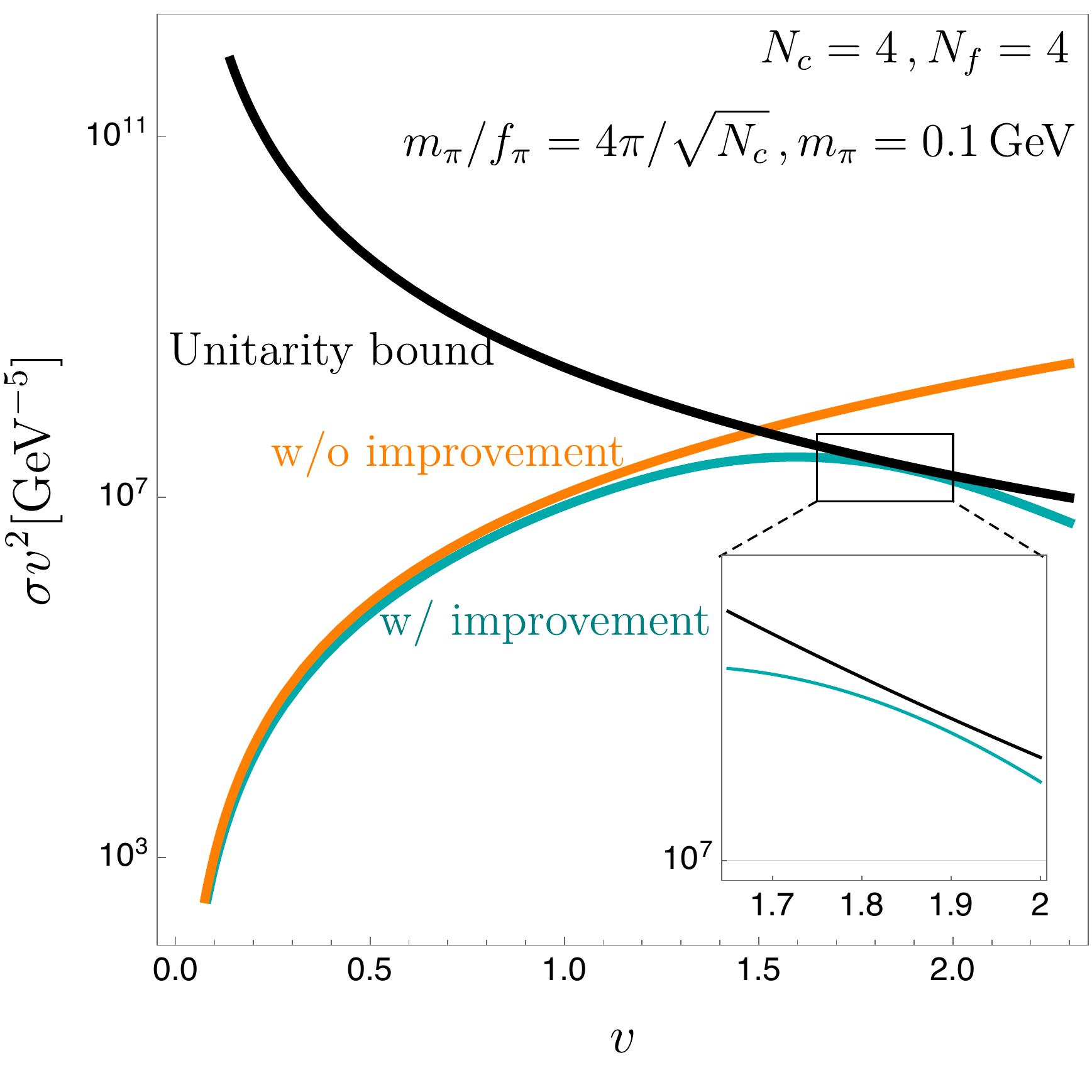}
	\includegraphics[width=0.45\linewidth]{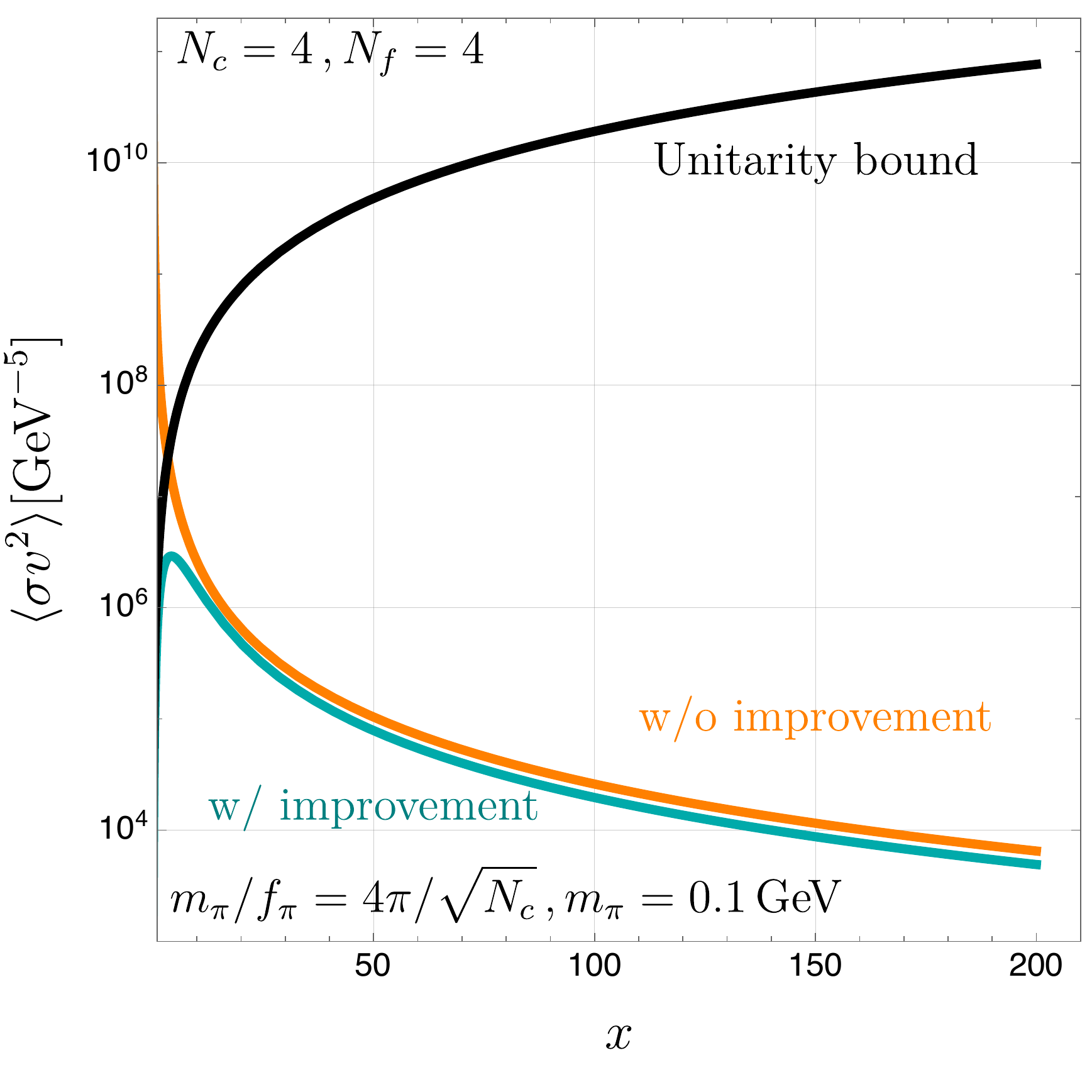}
	\caption{
		Comparison of the $\pi\pi\pi \to \pi \pi$ (thermally-averaged)  cross sections.
    The symmetry structure is assumed to be $SU(4) \times SU(4)/SU(4)$, and we take $m_\pi = 0.1 \,\mathrm{GeV}$.
    We take $m_\pi/f_\pi = 4 \pi$ in top panels and $m_\pi/f_\pi = 4 \pi/\sqrt{N_c}$ in bottom panels.
    (\textit{Left}): Shown is the cross sections $\sigma v^2$ with improvement (cyan), the cross section without improvement (orange), and unitarity bound (black) as a function of velocity $v$.
    (\textit{Right}): Shown is the thermally-averaged cross section $\langle\sigma v^2\rangle$ with improvement (cyan), without improvement (orange), and unitarity bound (black) as a function of $x = m_\pi/T$.
	}
	\label{fig:comparison}
\end{figure}

We compare the (thermally-averaged) cross section with improvement (cyan), that without improvement (orange), and the unitarity bound (black) in \cref{fig:comparison}.
We choose different $m_\pi/f_\pi$ for the top panels and for the bottom panels.
We use the improved partial-wave amplitude squared $|T^\mathrm{imp}_\mathrm{in}|^2$ in the $\pi\pi\pi \to \pi\pi$ cross section given by \cref{eq:crosssection_3to2}. 
In the left panels, we show the cross section $\sigma_{3 \to 2} v^2$ as the function of velocity $v = \overline k/\overline \mu$.
Since the tree-level partial-wave amplitude squared $(T^\mathrm{tree}_\mathrm{in})^2$ is proportional to $v^8$, the improved cross section follows the original cross section at small $v$. 
The improved cross section gets larger as $v$ increases, and the cross section almost saturates the unitarity bound at a certain $v$.
The partial-wave amplitude for each representation does not simultaneously saturate the unitarity bound at a certain $v$.
As we see in small windows in \cref{fig:comparison}, the improved cross section does not fully coincide with the unitarity bound, but approaches to the bound.
For a larger $v$, the improved partial-wave amplitude $(T^\mathrm{tree}_\mathrm{in})^2$ follows the scaling of $v^{-8}$, and hence the improved cross section is suppressed compared to the unitarity-saturated cross section, whose partial-wave amplitude follows the scaling of $v^{0}$.

The thermally-averaged cross section with the improvement hardly saturates the unitarity bound as shown in the right panels of \cref{fig:comparison} since $\sigma v^2$ with the improvement saturates the unitarity bound just in a narrow range of $v$.
A larger $v$ is required for saturating the unitarity bound for a smaller $m_\pi/f_\pi$, and hence the improvement hardly affects the thermally-averaged cross section for a smaller $m_\pi/f_\pi$ due to the Boltzmann suppression factor.
In particular, the tree-level inelastic amplitude squared is quite suppressed at small $v$ in the dark pion realizations, which is proportional to $v^8$, and hence the improvement has little impact on the cross section for the small $v$.
This is the main reasons why the improvement is less effective even for the coupling close to the na\"ive perturbative bound $m_\pi/f_\pi \simeq 4 \pi/\sqrt{N_c}$.
We note that there is a small difference between (thermally-averaged) cross sections with and without improvement even at small $v$ (large $x$).
The relative velocity of the two-body final state is given by $2 \beta \simeq 2 \sqrt{5}/3$ in the small $v$ limit, and hence the improvement by the $2 \to 2$ amplitude remains.

\begin{figure}
	\centering
	\includegraphics[width=0.32\linewidth]{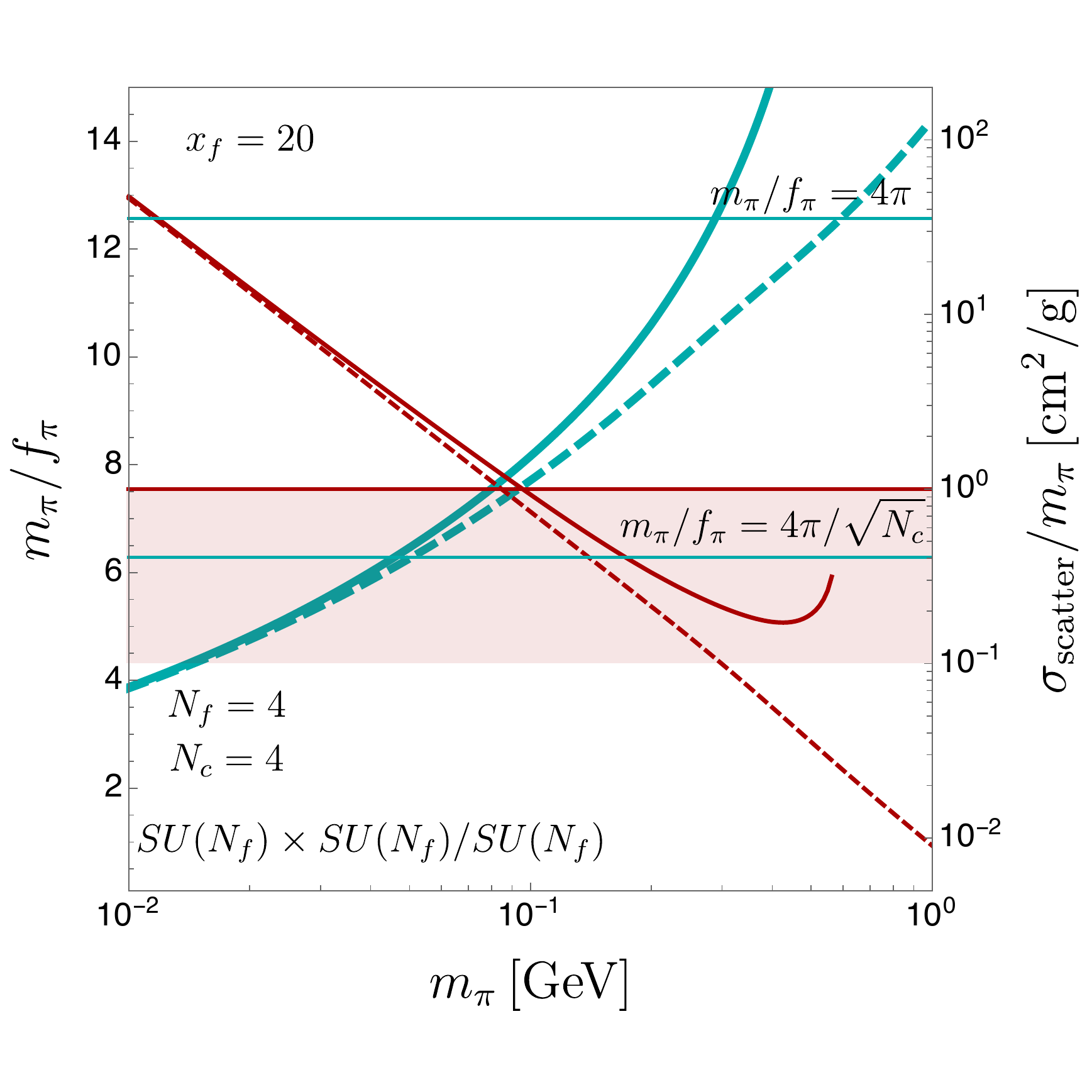}
	\includegraphics[width=0.32\linewidth]{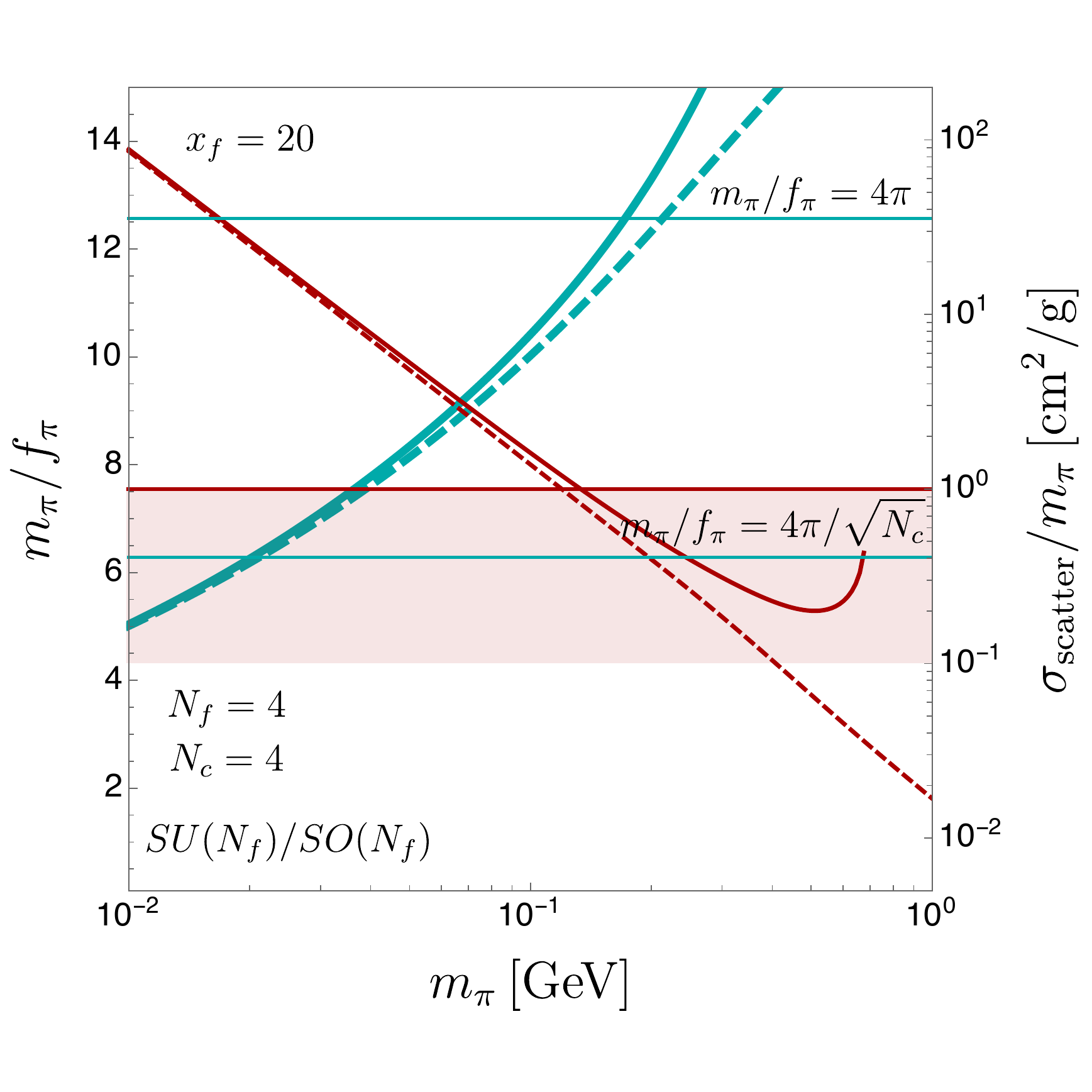}
	\includegraphics[width=0.32\linewidth]{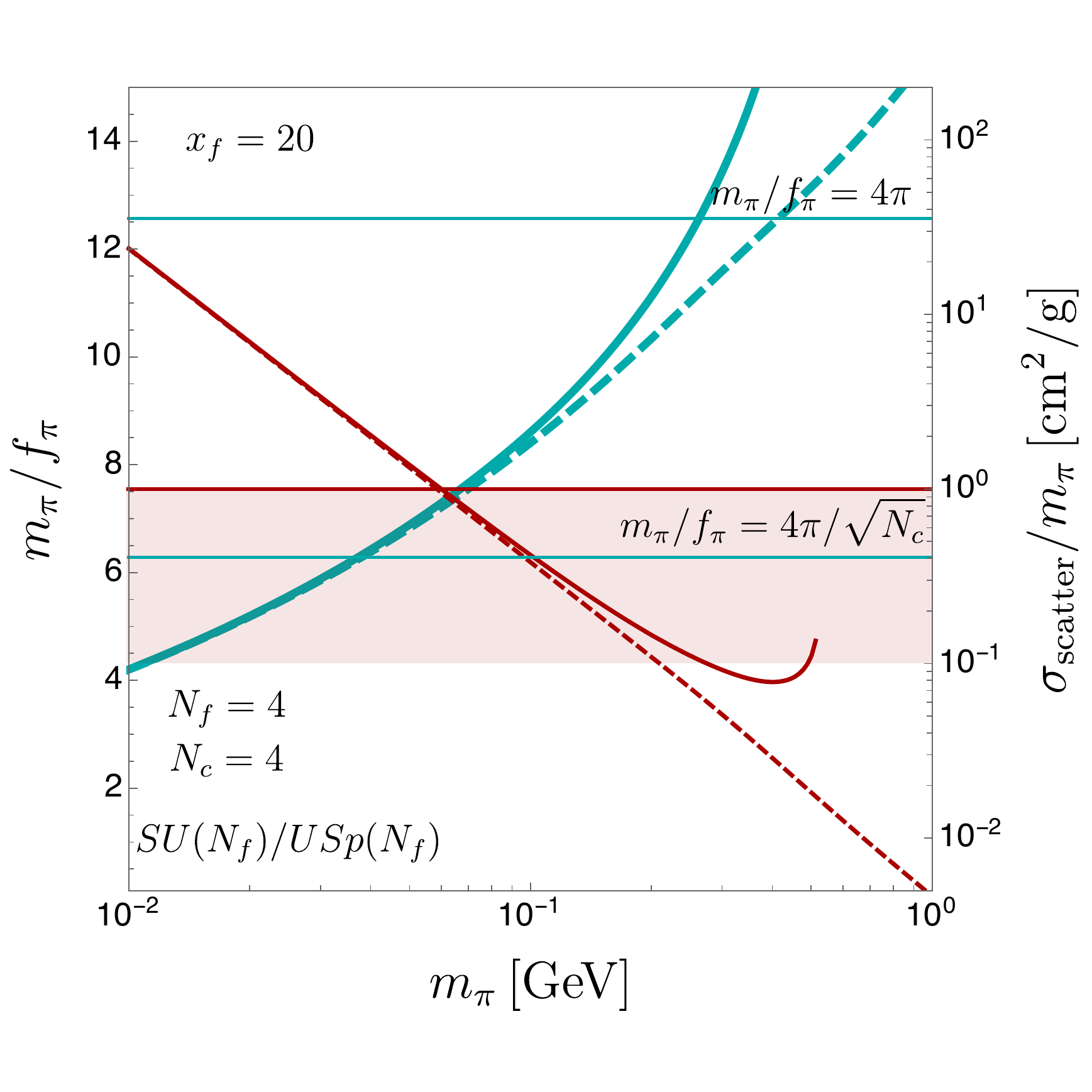} \\
	\includegraphics[width=0.32\linewidth]{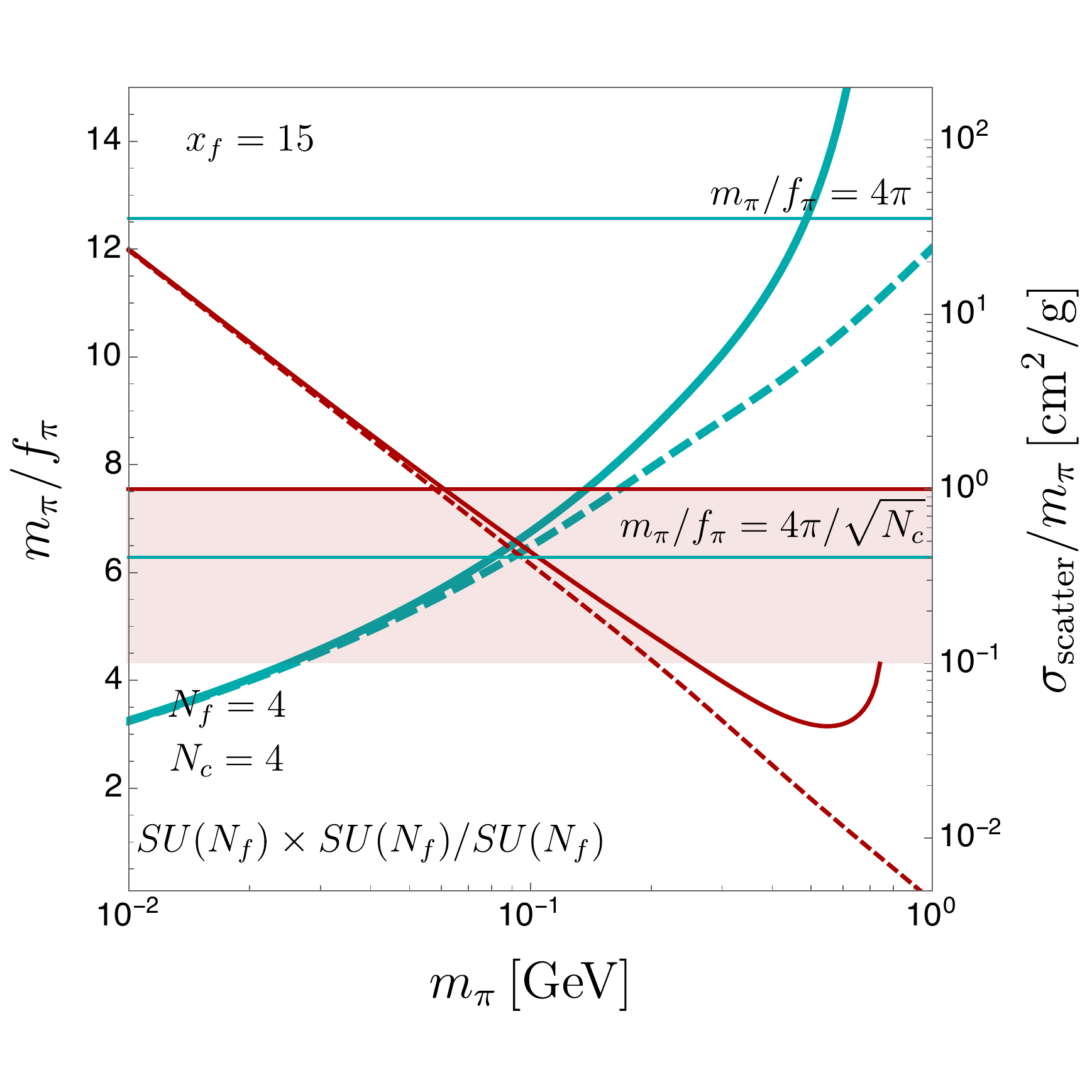} 
	\includegraphics[width=0.32\linewidth]{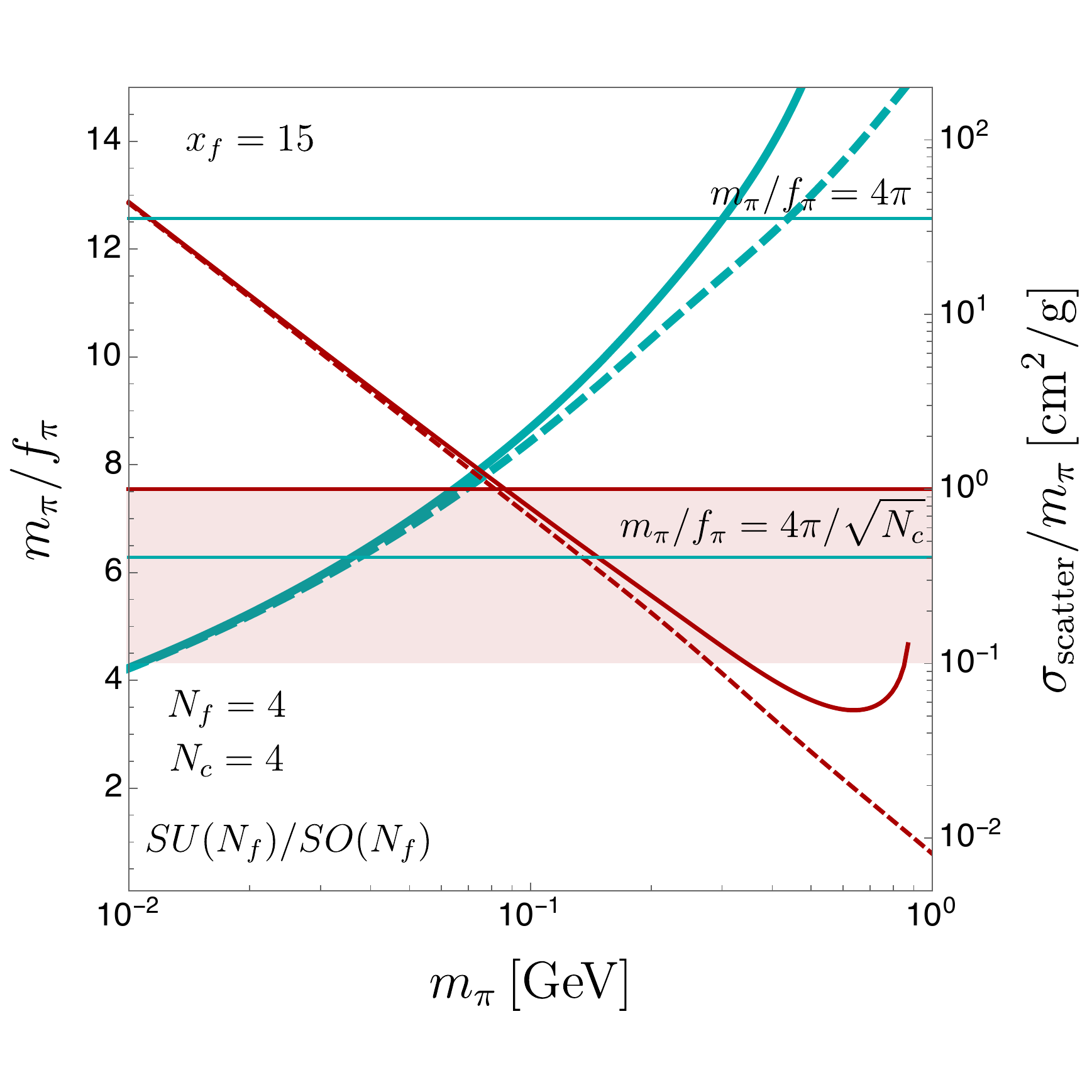} 
	\includegraphics[width=0.32\linewidth]{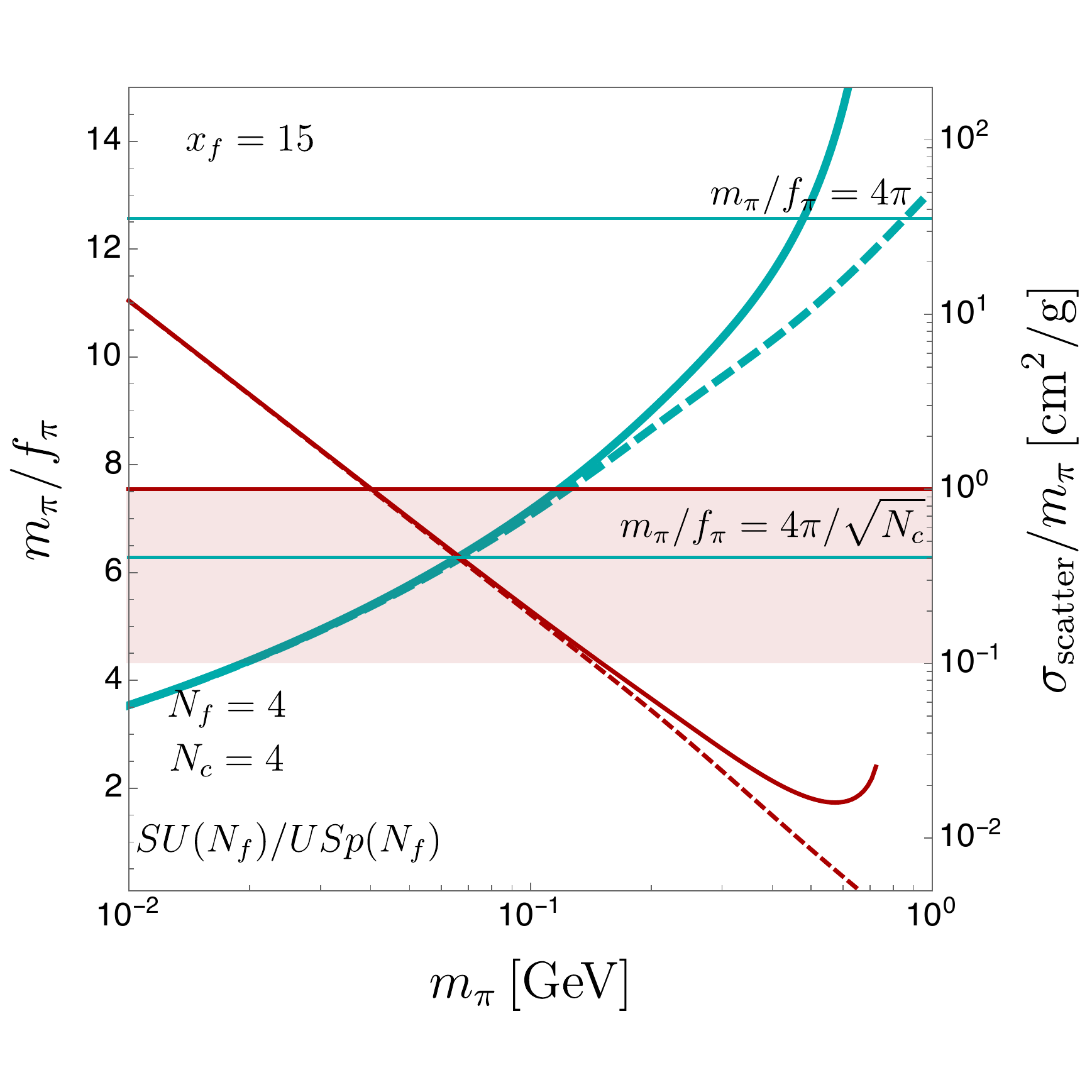}
	\caption{
		Shown are $m_\pi$-$m_\pi/f_\pi$ plots. 
    We assume different symmetry structures for each panel: 
    (left) $SU(N_f) \times SU(N_f)/SU(N_f)$, 
    (middle) $SU(N_f)/SO(N_f)$, and
    (right) $SU(N_f)/USp(N_f)$.
    We assume $N_f = 4$ and $N_c = 4$. 
    We take the different $x_f$: 
    (top) $x_f = 20$ and (bottom) $x_f =15$.
    The thick-cyan lines show required $m_\pi/f_\pi$ for given $m_\pi$ (values shown on the left-vertical axis): the thick-solid lines show the result using the improved cross section, while the thick-dashed lines show the result without improvement. 
    The thin-red lines show the self-scattering cross section with $m_\pi/f_\pi$ on thick lines (values shown on the right-vertical axis), and each line-type corresponds to that of thick lines.
    The red-shaded bands correspond to $0.1 \, \mathrm{cm^2/g} \leq \sigma_\mathrm{scatter}/m_\pi \leq 1 \, \mathrm{cm^2/g}$.
    The horizontal solid-lines show the unitary bound.
	}
	\label{fig:results}
\end{figure}

We illustrate the impact of the improvement of the $\pi \pi \pi \to \pi \pi$ cross section in \cref{fig:results}.
We compute the relic abundance of the SIMP DM by the use of an approximate formula, which ignores the temperature change of the effective degrees of freedom $g_\ast$ during freeze-out, given by Ref.~\cite{Choi:2017mkk}:
\eqs{
  \Omega h^2 \simeq \frac{1.05 \times 10^{-10} \, \mathrm{GeV}^{-2}}{g_{\ast}^{3/4}(T_f) m_\pi [K(x_f) / M_\mathrm{Pl}]^2} \,, \qquad 
  K(x_f) = \int_{x_f}^{\infty} \frac{dx}{x^5} \langle \sigma_{3 \to 2} v^2 \rangle \,.
}
Here, $M_\mathrm{Pl} = 2.4 \times 10^{18} \,\mathrm{GeV}$. 
We take $x_f = m_\pi/T_f = 15$ and $20$ in the figure.
We use the effective degrees of freedom $g_\ast$ at the freeze-out temperature given by Ref.~\cite{Saikawa:2020swg}. 

In \cref{fig:results}, we compare the required $m_\pi/f_\pi$ for the correct relic abundance $\Omega h^2 \simeq 0.12$ with the thermally-averaged cross section with and without the improvement (cyan lines): solid-thick lines for the improved cross section, while dashed-thick lines for the cross section without improvement.
The self-scattering cross section $\sigma_\mathrm{scatter}/m_\pi$, obtained along with $m_\pi/f_\pi$ required for the correct relic abundance, is shown as thin-red lines; these line-types correspond to that of cyan lines, namely, solid (dashed) lines for the cross section with (without) improvement.
We take the number of flavor and colors to be fixed as $N_f = 4$ and $N_c = 4$ in all panels.
Each panel shows the different symmetry structure. 
The horizontal solid-line shows the unitarity bounds $m_\pi/f_\pi = 4 \pi$ and $m_\pi/f_\pi = 4 \pi/\sqrt{N_c}$ given by \cref{eq:cutoffscale}, and the red-shaded band indicates the self-scattering cross section in a range of $0.1 \, \mathrm{cm^2/g} \leq \sigma_\mathrm{scatter}/m_\pi \leq 1 \, \mathrm{cm^2/g}$.

The required $m_\pi/f_\pi$ gets larger in order to explain the correct relic abundance as the pion mass gets increased.
The number-changing cross section is suppressed by the improvement procedure compared to the tree-level cross section for a large $m_\pi/f_\pi$, and hence the required $m_\pi/f_\pi$ is larger than the case without the improvement. 
We find that any large value of $m_\pi/f_\pi$ cannot explain the correct relic abundance above $m_\pi \simeq 700 \,\mathrm{MeV}$ ($m_\pi \simeq 900 \,\mathrm{MeV}$) for $x_f = 20$ ($x_f = 15$).
However, we have to incorporate the higher-order terms of the $\chi$PT when the pion self-coupling gets close to the na\"ive perturbative bound.

Once we take the na\"ive perturbative bound $m_\pi/f_\pi \simeq 4 \pi/\sqrt{N_c}$, there is no room for the allowed range of $m_\pi$ for $x_f = 20$ while there remains the allowed range in the left-bottom and right bottom panels for $x_f = 15$.
The improvement is less effective for the coupling close to the na\"ive perturbative bound $m_\pi/f_\pi \simeq 4 \pi/\sqrt{N_c}$.
Meanwhile, this bound gets mild when we take a weaker bound, e.g. $m_\pi/f_\pi \lesssim 4\pi$, and then the allowed range of $m_\pi$ can be enlarged.
As approaching to $m_\pi/f_\pi \simeq 4 \pi$, the improvement numerically modifies $m_\pi/f_\pi$ at the ten-percent level.

\section{Conclusion and Discussions \label{sec:conclusion}}

The dark pion is a promising realization of the SIMP models.
The pion self-coupling $m_\pi/f_\pi$ tends to be large in order that the SIMP framework works, and hence, it is expected that scattering cross sections easily violate the perturbative unitarity.
In this study, we have proposed the improvement of partial-wave amplitudes, which automatically hold the optical theorem. 
We have found that the impact of the improvement on phenomenology is less significant when the DM velocity is small.
Therefore, the improvement of the cross section does not change the self-interaction of dark pions at the cosmic structures.
Meanwhile, we have also examined the impact of the improvement on $\pi \pi \pi \to \pi \pi$ processes, whose final states are semi-relativistic.
The pion self-coupling required to achieve the correct relic abundance is hardly changed below the na\"ive perturbative bound $m_\pi/f_\pi \simeq 4 \pi/\sqrt{N_c}$ by the improvement.
The change will be more efficient near the original NDA limit $m_\pi/f_\pi \simeq 4 \pi$ \cite{Manohar:1983md}, and it can be up to $\mathcal{O}(10)\,\%$.
The impact of the improvement will be comparable to that of the NNLO correction of the $\chi$PT studied in Ref.~\cite{Hansen:2015yaa}, which is of order of $10\,\%$.
$m_\pi/f_\pi$ required for the correct relic abundance drastically changes for $m_\pi \gtrsim 300\,\mathrm{MeV}$, and we cannot find $m_\pi/f_\pi$ explaining the abundance for $m_\pi \gtrsim 700\,\mathrm{MeV}$. 
Since $m_\pi/f_\pi$ is beyond the na\"ive perturbative bound in the mass range, there are uncertainties due to the resonances and the higher-order corrections of the $\chi$PT. 

Ref.~\cite{Hansen:2015yaa} has discussed the impacts of the NLO and NNLO contributions of chiral Lagrangian.
They include only the real part of the amplitude since the imaginary part provides only the higher-order contribution to the cross section.
On the other hand, their amplitude does not satisfy the optical theorem.
Our improved amplitudes incorporate the imaginary part, but consists of the LO amplitudes of the elastic and inelastic scattering processes.
We may include the higher-order contributions systematically by using the inverse amplitude method, which automatically satisfies the optical theorem, in the absence of the inelastic channel~\cite{Dobado:1996ps}.
However, it is not straightforward to include inelastic channels in the method. 
This is beyond scope of this study.

In this study, we have focused only on the dark pion realization of the SIMP scenario. 
The amplitude for the number-changing process is suppressed in the following ways in the dark pion realization. 
The WZW term arises as the NLO contributions with a one-loop suppression factor in the chiral perturbation. 
Furthermore, the square of the inelastic amplitude is suppressed by $v^8$ in the dark pion realization, and it is a sub-leading contribution even in the $p$-wave contributions.
In other realizations, such as the dark glueball realization~\cite{Soni:2016gzf,Forestell:2016qhc} and the vector SIMP model~\cite{Choi:2017zww}, the square of the inelastic amplitude can be leading-order contribution of the perturbation theory and does not suffer from further velocity suppression.
There, unitarizing the amplitude will be more important within the na\"ive perturbative range in other realizations.

\subsection*{Acknowledgement}
A. K. acknowledges partial support from Grant-in-Aid for Scientific Research from the Ministry of Education, Culture, Sports, Science, and Technology (MEXT), Japan, 18K13535 and 19H04609; from World Premier International Research Center Initiative (WPI), MEXT, Japan; from Norwegian Financial Mechanism for years 2014-2021, grant nr 2019/34/H/ST2/00707; and from National Science Centre, Poland, grant 2017/26/E/ST2/00135 and DEC-2018/31/B/ST2/02283.
This work is also supported by the Advanced Leading Graduate Course for Photon Science and the JSPS Research Fellowships for Young Scientists (S.K.).
The work of T.K. is supported in part by the National Science Foundation of China under Grant Nos. 11675002, 11635001, 11725520, and 12235001.

\appendix

\section{One-loop amplitude in \texorpdfstring{$\lambda \phi^4$}{phi4} theory \label{app:phi4}}

Let us consider a $\lambda \phi^4$ theory of a scalar field $\phi$ with a mass of $m$. 
The tree-level amplitude for $\phi \phi \to \phi \phi$ is given by $i \mathcal{M}^\text{tree} = - i \lambda$, while the one-loop amplitude is given by
\eqs{
  i \mathcal{M}^\text{1-loop}
  & = - \frac{i \lambda^2}{32\pi^2} \int^1_0 dx 
  \left[
    \ln \frac{m^2 - x(1-x) s - i \epsilon}{\mu^2}
  \right. \\
  & \qquad \left. 
    + \ln \frac{m^2 - x(1-x) t - i \epsilon}{\mu^2}
    + \ln \frac{m^2 - x(1-x) u - i \epsilon}{\mu^2}
  \right] \,.
  \label{eq:phi1loop}
}
Here, $x$ denotes the Feynman parameter and $\mu$ denotes the renormalization scale. 
The Mandelstam variables $s\,,t\,,$ and $u$ are given in the center-of-mass frame by 
\eqs{
  s = 4 m^2 (1 + \beta^2 \gamma^2) \,, \quad
  t = - 2m^2 \beta^2 \gamma^2 (1 - \cos\theta) \,, \quad
  u = - 2m^2 \beta^2 \gamma^2 (1 + \cos\theta) \,, 
}
where $\beta^2 = 1 - 4 m^2/s$ and $\gamma^{-2} = 1-\beta^2$.
A branch cut appears along $\mathrm{Re}(s) > 4 m^2$.

We obtain the partial wave amplitudes by multiplying the Legendre polynomial and integrating over the angle $\theta$.
We focus only on the $s$-wave partial amplitude. 
The tree-level amplitude is given by
\eqs{
  i T_0^\text{tree} \equiv 
  \frac{1}{64 \pi} \int^1_{-1} d\cos\theta (i \mathcal{M}^\text{tree}) = - \frac{i \lambda}{32 \pi} \,,
}
while the one-loop amplitude is 
\eqs{
  i T_0^\text{1-loop} & \equiv 
  \frac{1}{64 \pi} \int^1_{-1} d\cos\theta (i \mathcal{M}^\text{1-loop}) \\ 
  & = \frac{i \lambda^2}{512\pi^3} 
  - \frac{i \lambda^2}{1024\pi^3} \int^1_0 dx 
  \left[
    \ln \frac{m^2 - x(1-x) s - i \epsilon}{\mu^2}
  \right. \\
  & \qquad \left. 
    + 2 \ln \frac{m^2}{\mu^2}
    +  \frac{1 + 4 x(1-x) \beta^2 \gamma^2}{2 x(1-x) \beta^2 \gamma^2} \ln [1 + 4 x(1-x) \beta^2 \gamma^2]
  \right] \,.
}
Here, we follow the definition of the partial-wave amplitude \cref{eq:pwdecomp}.
The one-loop corrected amplitude \cref{eq:phi1loop} is totally symmetric under the exchange of $s\,, t\,, $ and $u$. 
Therefore, the partial-wave amplitude has another branch cut, which corresponds to branch cuts along $\mathrm{Re}(t) > 4 m^2$ or $\mathrm{Re}(u) > 4 m^2$. 
In this region, $\mathrm{Re}(s) < 0$, and hence $\beta^2 \gamma^2 < 0$. 
The another branch cut along $\mathrm{Re}(s) < 0$ originates from the last term. 

As far as we consider the physical region of the scattering process, $\mathrm{Re}(s) > 0$, the imaginary part of $T_0^\text{1-loop}$ appears only from the first logarithmic term. 
The argument inside the logarithmic function is negative only when $(1 - \beta)/2 < x < (1 + \beta)/2$, and hence the imaginary part of the one-loop partial-wave amplitude is 
\eqs{
  \mathrm{Im} (T_0^\text{1-loop}) & = 
  - \frac{\lambda^2}{1024\pi^3} \int^1_0 dx 
  \mathrm{Im} 
  \left[
    \ln \frac{m^2 - x(1-x) s - i \epsilon}{\mu^2}
  \right] \\
  & = \frac{\lambda^2}{1024\pi^2} \sqrt{1 - \frac{4m^2}{s}} 
  = \beta (T_0^\text{tree})^2 \,.
}
Here, we confirm the optical theorem \cref{eq:opticaltheorem} without the inelastic scattering at the LO of the coupling $\lambda$ in the $\lambda \phi^4$ theory. 

Finally, we discuss the coupling expansion of the total cross section and the importance of the imaginary part.
We find the total cross section up to the one-loop level as follows:
\eqs{
  \sigma_\mathrm{total} & = \frac{32\pi}{s} |T_0^\text{tree}+T_0^\text{1-loop}|^2 \\
  & \simeq \frac{\lambda^2}{32\pi s} \left[ 1 +\frac{2 \lambda}{32\pi^2} + \beta^2 \left( \frac{\lambda}{32\pi} \right)^2 + \left( \frac{\lambda}{32\pi^2} \right)^2 \right] \,.
}
Here, we ignore the logarithmic factors of the real part of the one-loop correction in this expression.
In the square bracket, the first term corresponds to the tree-level contribution, the second is the interference between the tree-level and the real part of the one-loop correction, the third shows the imaginary part of the one-loop correction, and the last is the contribution from the real part of the one-loop amplitude. 
We note that the last-two terms must be affected by the interference between the tree-level and 2-loop amplitudes. 
All loop corrections similarly contribute to the cross section when $16 \pi^2 \lesssim \lambda$, and hence the coupling expansion is no longer validated.
The improved amplitude in the text incorporates only the imaginary part, and hence the improvement is expected to be important as the coupling is just below the na\"ive perturbative bound. 

\section{N-body Cross Sections and Thermal Averaging \label{app:crosssection}}
In this appendix, we discuss the $N$-body cross section and its thermal averaging. 

\subsection{Cross sections of non-relativistic $N$-body system}

We consider the $N$-body system, which is composed of $N$ particles with momentum $\vec{p}_i$ and mass $m_i$ ($i = 1, \cdots, N$).
Jacobi coordinates of the system are often useful to simplify the $N$-body system. 
When the particle $i$ is positioned at $\vec{x}_i$, the Jacobi coordinates of the system are defined by
\eqs{
  \vec{r}_i & = \frac{1}{m_1 + \cdots + m_i} \left[ \sum_{j=1}^i m_{j} \vec{x}_{j} - \sum_{j=1}^i m_{j} \vec{x}_{i+1}\right] \,, \; (i = 1, \cdots, N-1) \,, \\
  \vec{r}_N & = \frac{1}{m_1 + \cdots + m_N} \sum_{j=1}^N m_{j} \vec{x}_{j} \,.
} 
Here, $\vec{r}_i \, (i = 1, \cdots, N-1)$ represents the relative coordinates between the position of the particle $i+1$ and the center-of-mass position for the sub-system of $i$ particles.
$\vec{r}_N$ denotes the center-of-mass position of the total system.
We introduce Jacobi coordinates of the momentum space, which we call the Jacobi momenta $\vec{k}_i$ $(i = 1, \cdots, N-1)$ and the center-of-mass momentum $\vec{k}_N$.
\eqs{
  \vec{k}_i & = \frac{1}{m_1 + \cdots + m_{i+1}} \left[ m_{i+1} \sum_{j=1}^i \vec{p}_{j} - \sum_{j=1}^i m_{j} \vec{p}_{i+1}\right] \,, \; (i = 1, \cdots, N-1) \,, \\
  \vec{k}_N & = \vec{p}_{1} + \cdots + \vec{p}_{N} \,.
} 
The Jacobi momenta corresponds to the relative momentum of a particle $i+1$ and the center-of-mass of subsystem with $i$ particles.
The total angular momentum is written as the vector product of the Jacobi coordinates and Jacobi momenta.
\eqs{
  \vec{L} \equiv \sum_{i=1}^N \vec{x}_i \times \vec{p}_i = \sum_{i=1}^N \vec{r}_i \times \vec{k}_i \,.
} 
In terms of $\vec{k}_i$ $(i = 1, \cdots, N)$, the total energy of the system is 
\eqs{
  \sum_{i} \frac{\vec{p}_i^2}{2 m_i} = \sum_{i} \frac{\vec{k}_i^2}{2 \mu_i}  \,.
}
Here, $\mu_i \, (i = 1, \cdots, N-1)$ denotes the reduced mass of a particle $i+1$ and the center-of-mass of subsystem with $i$ particles, 
\eqs{
  \frac{1}{\mu_i} = \frac{1}{m_1 + \cdots + m_i} + \frac{1}{m_{i+1}} \,, \quad (i = 1, \cdots, N-1) \,,
}
and $\mu_N$ denotes the total mass of all particles, $\mu_N =  m_1 + \cdots + m_N$. 

We compute the phase space integral by the use of the Jacobi momenta. 
The Jacobian of this transformation is unity, and hence the $N$-body phase space integral is computed as follows. 
\eqs{
  d \Phi_N & \equiv \prod_{i=1}^N \frac{d^3 p_i}{(2\pi)^3 2 m_i} (2 \pi)^4 \delta^{(4)}(P - \sum p_i) \\
  & = \prod_{i=1}^N \frac{d^3 k_i}{(2\pi)^3 2 m_i} (2 \pi)^4 \delta(E - \sum \frac{k_i^2}{2 \mu_i})\delta^{(3)}(\vec{P} - \vec{k}_N) \,.
}
Here, $P$ denotes the total 4-momentum $P = (E, \vec{P})$.
One of advantages in using the Jacobi momenta is the simplification of the momentum integral with the delta function.
The delta function of 3-momentum is given as a function of a single variable $\vec{k}_N$ by the use of the Jacobi momenta, and hence we can easily carry out the $k_N$-integral.
We take the center-of-mass frame of $\vec{p}_i$, with $\vec{P} = \vec{0}$, in the following.
We can simplify the phase space integral by replacing the variables $\vec{\kappa}_i = \sqrt{\overline \mu_N/\mu_i} \vec{k}_i$ with 
\eqs{
  (\overline \mu_N)^{N-1} = \mu_1 \cdots \mu_{N-1}
  = \frac{m_1 \cdots m_N}{m_1 + \cdots + m_N} \,.
}
The phase space integral takes the following form:
\eqs{
  d \Phi_N 
  & = \prod_{i=1}^N \left(\frac{1}{2 m_i} \right) \frac{d \overline k_N d\Omega_{3N-3}}{(2\pi)^{3N-4}} \overline k_N^{3N-4}  \delta\left(E - \frac{\overline k_N^2}{2\overline \mu_N} \right) \,.
}
Here, $d\Omega_d$ denotes the integration over the solid angle in $d$ dimensions, and $\overline k_N^2 \equiv \kappa_1^2 + \cdots + \kappa_{N-1}^2$.
The solid angle in $d$ dimensions is given by $\Omega_d = 2 \pi^{d/2}[\Gamma(d/2)]^{-1}$.
We note that the subscripts for $\overline \mu_N$ and $\overline k_N$ just denote the $N$-body phase space, not $N$-th variables. 

Now, we consider 2-body elastic and inelastic cross sections of identical scalar particles $\chi_i$.
The scattering amplitudes for these processes are generally the functions of solid angles for the initial states and final states, even though these are independent of the angles for the 2-body scattering.
Integrating the amplitude over the angle of the final state, and averaging over the angle of the initial state, we obtain the averaged cross sections~\cite{Mehta:2009}.
For our purpose, we consider the $2 \to 2$ elastic scattering and the $2 \to N$ inelastic scattering.
These cross sections take the form
\begin{align}
  \sigma_{2 \to 2} 
  & = \frac{2!}{\Omega_3} \left( \frac{2\pi}{p} \right)^2 \frac{1}{N_\chi^2} \sum_R \sum_\ell (2 \ell + 1) \times 4 \beta^2 |T^R_\ell|^2 d_R \,, \\
  \sigma_{2 \to N} 
  & = \frac{2!}{\Omega_3} \left( \frac{2\pi}{p} \right)^2 \frac{1}{N_\chi^2} \sum_R \sum_\ell (2 \ell + 1) \times 4 \beta^2 |T_{\mathrm{in},\ell}^R|^2 d_R \,. \label{eq:nonrel_2toN} 
\end{align}
Here, $T^R_\ell$ and $|T_{\mathrm{in},\ell}^R|^2$ are the partial-wave amplitude integrated over the phase space of the initial state.
$N_\chi$ denotes the number of degrees of freedom of $\chi_i$.
$p = \sqrt{s} \beta/2$ is the momentum of incoming particles. 
$\beta = (1-4m^2/s)^2$ for the identical (up to flavor indices) particles with the mass $m$, where $\sqrt{s}$ is the collision energy.
We use the relativistic formula for the two-body system ($p$ and $\beta$), while taking the non-relativistic limit of the $N$-body system. This is because the two-body system is semi-relativistic for the $2 \to N$ process.
The prefactor in the cross section originates from the two-body phase space integral including a factor for the identical particles, $2!$.
A factor of $4 \beta^2$ in front of the partial-wave amplitude originates from the definition of the $S$-matrix elements $S_\ell$ for the partial-wave $\ell$, $S_\ell = 1 + 2 i \beta T_\ell$.

Once the system has the time-reversal invariance, the partial-wave amplitudes for the $2 \to N$ scattering and for the $N \to 2$ scattering are given by $|T_{\mathrm{in},\ell}^R|^2$. 
In a similar way to the 2-body scattering, we define the initial-angle averaged scattering cross section for $N \to 2$ process in the non-relativistic limit of the initial states.
\eqs{
  \sigma_{N \to 2} v^{N-2}
  & = \frac{N!}{\Omega_{3N-3}} \left( \frac{2\pi}{\overline k_N} \right)^{3N-4} \frac{1}{N_\chi^N} \sum_R \sum_\ell (2 \ell + 1) \times 4 \beta^2 |T_{\mathrm{in},\ell}^R|^2 d_R \,. 
  \label{eq:nonrel_Nto2}
}
Here, the prefactor is due to averaging the $N$ initial states. 
We may define the velocity for the $N$-body states as follows: $v \equiv \overline k_N/ \overline \mu_N$.

Now, we compute the thermally-averaged cross section. 
Another advantage of the use of the Jacobi momenta is that the Jacobi momenta are invariant under the Galilei transformation except for $\vec{k}_N$. 
Indeed, the momentum variables of the thermal averaging do not necessarily coincide with those of a scattering system, and hence the difference of the momentum variables \textit{does} affect the thermal averaging unless we consider only $s$-wave process.
The Jacobi momenta describe the relative motion of the subsystem except for $\vec{k}_N$ describing the motion of the center-of-mass.
The cross section must be expressed only by the Jacobi momenta (except for $\vec{k}_N$) since the cross section depends only on the relative motions of scattering particles.
The thermally-averaged cross section is defined as
\eqs{
  \langle \sigma_{N \to 2} v^{N-1} \rangle
  & \equiv \prod_{i=1}^N \frac{1}{n_i} \int \frac{d^3 p_i}{(2\pi)^3 2 m_i} e^{- \frac{p_i^2}{ 2 m_i T}} (\sigma_{N \to 2} v^{N-1}) \,, \\
  n_i 
  & \equiv \int \frac{d^3 p_i}{(2\pi)^3 2 m_i} e^{- \frac{p_i^2}{ 2 m_i T}} \,. 
}
We substitute the integral variables $\vec{p}_i \to \vec{k}_i$, and the Jacobian is unity as stated before.
The integral of the cross section with respect to $\vec{k}_N$ is trivially carried out since the cross section does not depend on $\vec{k}_N$. 
We may simplify the integral again by replacing the variables $\vec{k}_i \, (i = 1, \cdots, N-1)$ to $\overline k_N$ and hyperangles as shown before, and the phase space integral of the initial state may be rewritten as 
\eqs{
  \langle \sigma_{N \to 2} v^{N-1} \rangle
  & = \frac{1}{n} \prod_{i=1}^N \left(\frac{1}{2 m_i} \right) \int \frac{d \overline k_N d\Omega_{3N-3}}{(2\pi)^{3N-4}} \overline k_N^{3N-4} e^{- \frac{\overline k_N^2}{ 2 \overline \mu_N T}} (\sigma_{N \to 2} v^{N-1}) \,, \\
  n
  & \equiv \prod_{i=1}^N \left(\frac{1}{2 m_i} \right) \int \frac{d \overline k_N d\Omega_{3N-3}}{(2\pi)^{3N-4}} \overline k_N^{3N-4} e^{- \frac{\overline k_N^2}{ 2 \overline \mu_N T}} \,. 
  \label{eq:thermal_Nto2}
}
Here, averaging over the angle variables gives the initial-angle averaged cross section (\ref{eq:nonrel_Nto2}).
Taking the partial-wave amplitudes to be the maximum value $|T_{\mathrm{in},\ell}^R|^2 \leq (1/2\beta)^2$, we obtain the upper bound of the thermally-averaged cross section for $N \to 2$ process.
\eqs{
  \langle \sigma_{N \to 2} v^{N-1} \rangle_\mathrm{uni}
  = \frac{\left(2 \pi x \right)^{\frac{3N-5}{2}}}{m_\chi^{3N-4}} \frac{N\sqrt{N}}{N_\chi^N} \sum_R \sum_\ell (2 \ell + 1) d_R \,.
}
Here, $R$ and $\ell$ run only over representations whose amplitudes are non-zero, $x = m_\chi/T$, and we take the identical mass to be $m_\chi$.
We again note that we independently obtain the upper bounds from Refs.~\cite{Kuflik:2017iqs,Namjoo:2018oyn,Bhatia:2020itt}.
As we have stated in the text, our result is consistent with Refs.~\cite{Kuflik:2017iqs,Namjoo:2018oyn} up to a factor of $1/4$ due to maximizing the partial-wave amplitude, but our result disagrees with the results in Ref~\cite{Bhatia:2020itt} by a factor of $N\sqrt{N}/2\sqrt{2}$.

\subsection{Application to the SIMP models}

The tree-level five-pion amplitude is given by \cref{eq:5ptvertex} in the text.
We compute the thermally-averaged cross section of the $3 \to 2$ process. 
First, we compute the $2 \to 3$ cross section in the non-relativistic limit of the three-body final state, and then we obtain the $3 \to 2$ cross section by the use of the relation between cross sections \cref{eq:nonrel_2toN,eq:nonrel_Nto2} (with $N = 3$).
We take the center-of-mass frame of the initial pions of the $2 \to 3$ process, $\vec{p}_4+\vec{p}_5 = \vec{0}$, and the collision energy is assumed to be $\sqrt{s}$, and thus the 4-momentum of the initial pion is given by 
\eqs{
  p_4 = \frac{\sqrt{s}}{2} (1, \vec{\beta}) \,, \qquad 
  p_5 = \frac{\sqrt{s}}{2} (1, -\vec{\beta}) \,, 
}
with $|\vec{\beta}| = (1-4m_\pi^2/s)^{1/2}$.
The amplitude takes the form
\eqs{
  \mathcal{M} = - \frac{16 k}{\pi^2 f_\pi^5} T_{[abcde]} \frac{s}{2} \vec{\beta} \cdot (\vec{p}_1 \times \vec{p}_2) \,.
  \label{eq:3-to-2_amplitude}
}
We carry out the three-body phase-space integral for the final-state pions to obtain the $2 \to 3$ cross section. 
We introduce a practical technique to compute the six-dimensional hyperspherical integral.
$\overline k_3^3 = \kappa_1^2 + \kappa_2^2$ since we take the center-of-mass frame. 
We may parametrize each momentum as follows.
\eqs{
  \vec{\kappa}_1 & = \overline k_3 \cos \theta (\sin\theta_1 \cos\phi_1 \,, \sin\theta_1\sin\phi_1 \,, \cos\theta_1) \,, \\
  \vec{\kappa}_2 & = \overline k_3 \sin \theta (\sin\theta_2 \cos\phi_2 \,, \sin\theta_2\sin\phi_2 \,, \cos\theta_2) \,.
}
Here, $(\theta_i \,, \phi_i)$ denotes the solid angle for the momentum $\vec{\kappa}_i$, and $\theta$ distributes $\overline k_3$ to $\kappa_1$ and $\kappa_2$.
This assignment of angles is called as the Delves coordinate~\cite{delves1958tertiary,delves1960tertiary}.
The integral measure is 
\eqs{
  \int d \Omega_6 = \int^{\pi/2}_0 d \theta \sin^2 \theta \cos^2 \theta d\Omega^1_3 d\Omega^2_3 \,, 
}
where $d\Omega^i_3$ denotes the solid-angle integral for the momentum $\vec{\kappa}_i$.
The three-body phase space integral in the non-relativistic limit takes the form:
\eqs{
  d \Phi_3
  & = \prod_{i=1}^3 \left(\frac{1}{2 m_i} \right) \frac{1}{(2\pi)^{5}} \overline \mu_3 \overline k_3^{4} d \theta \sin^2 \theta \cos^2 \theta d\Omega^1_3 d\Omega^2_3 \,.
  \label{eq:NR_3bodyPS}
}
Here, $\overline k_3^2 = 2 \overline \mu_3 E$. 
Meanwhile, the relativistic form of the three-body phase space integral is given by
\eqs{
  d \Phi_3
  & = \frac{d s_{23}}{16} \frac{1}{(2\pi)^{5}} d\Omega^1_3 d\Omega^2_3 \overline \beta\left( \frac{m_1^2}{s}\,, \frac{s_{23}}{s} \right) \overline \beta\left( \frac{m_2^2}{s_{23}}\,, \frac{m_3^2}{s_{23}} \right) \,, \\
  \overline \beta\left( a , b \right) 
  & \equiv \frac12 \sqrt{1 - 2 (a+b) + (a-b)^2} \,.
}
where $s_{23} = (p_2+p_3)^2$.
The relativistic counterpart of the Delves angle $\theta$ is the Dalitz variable $s_{23}$.
$s_{23}$ takes a value within the range of $(m_2+m_3)^2 \leq s_{23} \leq (\sqrt{s}-m_1)^2$.
We may consider the non-relativistic limit of the integral by parameterizing as $\sqrt{s_{23}} = m_2+m_3 + E \cos^2 \theta$ with $E = \sqrt{s}-(m_1+m_2+m_3)$ being the center-of-mass energy in the limit.
We expand $\overline \beta$ and $d s_{23}$ in small $E$ limit as follows.
\eqs{
  \overline \beta\left( \frac{m_1^2}{s}\,, \frac{s_{23}}{s} \right) \overline \beta\left( \frac{m_2^2}{s_{23}}\,, \frac{m_3^2}{s_{23}} \right)
  & \simeq \sin\theta \cos\theta \frac{2 \overline \mu_3^3 E}{m_1m_2m_3(m_2+m_3)} \,, \\
  d s_{23} 
  & \simeq 4 E (m_2+m_3) d\theta \sin\theta \cos\theta \,,
}
and therefore, we obtain the non-relativistic form of the phase-space integral given by \cref{eq:NR_3bodyPS} by replacing $E = (\overline k_3^2/2\overline \mu_3)$.
The $2 \to 3$ cross section in the non-relativistic limit of the three-body final state is 
\eqs{
  \sigma_{2 \to 3} 
  & = \frac{1}{3!} \frac{1}{2 s \beta} \frac{1}{N_\pi^2} \int d \Phi_3 |\mathcal{M}|^2 
  =  \frac{1}{3!} \frac{1}{(2\pi)^5} \frac{k^2}{6 \pi} \frac{\overline k_3^8 \overline \mu_3 s \beta}{m_\pi^3 f_\pi^{10}} \frac{1}{N_\pi^2} \sum_R t_R^2 \,.
}
Here, $t_R^2$ is the decomposition of $t^2$ into each representations.
We carry out the three-body phase-space integral by following \cref{eq:NR_3bodyPS}.
By comparing the cross section with \cref{eq:nonrel_2toN} (with $N = 3$), we find the square of the partial-wave amplitude (\ref{eq:pw_inelastic}) for the inelastic scattering in the text.

The thermal averaging of the $3 \to 2$ annihilation is defined by \cref{eq:thermal_Nto2} (with $N = 3$). 
Since the partial-wave amplitude is averaged over hyperangles, the hyperangle integral is trivially carried out and only $\overline k_3$-integral remains. 
We find the thermally-averaged cross section as follows.
\eqs{
  \langle \sigma_{3 \to 2} v^2 \rangle = \frac{5 \sqrt5 k^2 m_\pi^5}{6 \pi^5 x^2 f_\pi^{10} } \frac{1}{N_\pi^3} \sum_R t_R^2 \,.
  \label{eq:thermal_averaged_3-to-2}
}
Our result is smaller than the cross section in the original paper~\cite{Hochberg:2014kqa} by a factor of $1/3$.
The difference could arise from the Galilei transformation of the invariant amplitude \cref{eq:3-to-2_amplitude}, which is computed in the center-of-mass frame.
$\vec{p}_1 \times \vec{p}_2$ is not Galilean invariant, and in fact it transforms as $\vec{p}_1 \times \vec{p}_2 \to \vec{p}_1 \times \vec{p}_2 + m_2 \vec{p}_1 \times \vec{v} + m_1 \vec{v} \times \vec{p}_2$ under $\vec{p}_i \to \vec{p}_i + m_i \vec{v}$ with $\vec{v}$ being a constant velocity.
The corresponding part may be rewritten in terms of the Jacobi momentum as follows.
\eqs{
  \vec{\beta} \cdot (\vec{p}_1 \times \vec{p}_2)
  = \vec{\beta} \cdot (\vec{k}_1 \times \vec{k}_2)
  + \frac23 \vec{\beta} \cdot (\vec{k}_1 \times \vec{k}_3) \,.
}
As we mentioned, $\vec{k}_3 = 0$ in the center-of-mass frame.
Hence, the second term vanishes since the amplitude (\ref{eq:3-to-2_amplitude}) is computed in the center-of-mass frame, and then we obtain \cref{eq:thermal_averaged_3-to-2}.
Meanwhile, we reproduce the result in Ref.~\cite{Hochberg:2014kqa} by keeping the second term artificially and by keeping $k_3$ integral with the Maxwell distribution (with $\mu_3$).
We note again that it is important to rewrite the cross section in terms of the Jacobi momenta, namely the Galilean-invariant way, especially for computing thermally-averaged cross section.

\section{Group-theoretical Computation \label{app:group}}

We consider the irreducible decomposition of products of broken generators of $G/H$. 
It is convenient to use the $SU(N_f)$ generators for describing the broken generators even for $G/H = SU(N_f)/SO(N_f)$ and $SU(N_f)/USp(N_f)$.
The fundamental generators are denoted by $T^a$, whose normalization is defined by
\eqs{
  \mathrm{tr} (T^a T^b) = \frac12 \delta^{ab} \,.
}
The Fierz identity for the generator is 
\eqs{
  (T^a)_i^{~j}(T^a)_k^{~l} = \frac12 \left( \delta_i^l \delta_k^j - \frac{1}{N_f} \delta_i^j \delta_k^l \right) \,.
}
The commutation and anticommutation of the generators gives the group-theoretical constants, $d$ and $f$ as 
\eqs{
  [T^a, T^b] = i f^{abc} T^c\,, \qquad
  \{T^a, T^b\} = d^{abc} T^c + \frac{1}{N_f} \delta^{ab}\,,
}
and we use these constants for the short-hand notation.

%%%%%%%%%%%%%%%%%%%%%%%%%%
\subsection{\texorpdfstring{$SU(N_f)\times SU(N_f)/SU(N_f)$}{SU(N) times SU(N)/SU(N)}}

First, let us consider $G = SU(N_f) \times SU(N_f)$ and $H = SU(N_f)$. 
The broken generator is the adjoint representation of $SU(N_f)$, and the direct product of the generator is symbolically given by 
\eqs{
  \mathrm{Ad} \otimes \mathrm{Ad}
  = 1_S 
  \oplus \mathrm{Ad}_A \oplus \mathrm{Ad}_S
  \oplus T_A \oplus \overline T_A 
  \oplus Y_S \oplus S_S \,.
}
Here, $\mathrm{Ad}$ denotes the adjoint representation of $SU(N_f)$.
The subscripts $S$ and $A$ denote symmetrization and anti-symmetrization of indices of adjoint representations in the left-hand side.
Other representations are summarized in \cref{tab:reps}.
As seen in the table, several representations vanish for $N_f \leq 3$.
For $N_f = 2$, $\mathrm{Ad}_S$, $T_A \oplus \overline T_A$, and $Y_S$ do not appear and the decomposition is given by
\eqs{
  \mathbf{3} \otimes \mathbf{3}
  = \mathbf{1}_S \oplus \mathbf{3}_A \oplus \mathbf{5}_S, 
}
and for $N_f = 3$, $Y_S$ does not appear and the decomposition is given by
\eqs{
  \mathbf{8} \otimes \mathbf{8}
  = \mathbf{1}_S \oplus \mathbf{8}_A \oplus \mathbf{8}_S \oplus \mathbf{10}_A \oplus \overline{\mathbf{10}}_A \oplus \mathbf{27}_S \,.
}

\begin{table}[t]
  \centering
	\caption{Representations in $SU(N_f)$ and their dimensions
	}
	\label{tab:reps}
  \renewcommand\arraystretch{2}
  \begin{tabular}{|c|c|}
    \hline
    Reps. & Dims. \\ \hline
    $\mathrm{Ad}$ & $N_f^2-1$ \\
    $T (\overline T)$ & $\dfrac14 (N_f^2-1)(N_f^2-4)$ \\
    $Y$ & $\dfrac14 (N_f+1) N_f^2 (N_f-3)$ \\
    $S$ & $\dfrac14 N_f^2 (N_f+3)(N_f-1)$ \\ \hline
  \end{tabular}
\end{table}

The projection operators with adjoint indices are defined by 
\eqs{
  P_R^{ab;cd} = 4 (T^a)_j^{~i} (T^b)_l^{~k} (P_R)^{jl;i'k'}_{ik;j'l'} (T^c)_{i'}^{~j'} (T^d)_{k'}^{~l'} \,.
}
Here, the subscript $R$ stands for the representations, and a factor of 4 is for the Dynkin normalization.
$(P_R)^{jl;i'k'}_{ik;j'l'}$ denotes the projection operators for the fundamental indices, which reflect the symmetries of corresponding Young tableaux. 
Now, we decompose the product $(T^a)_j^{~i} (T^b)_l^{~k}$ into its irreducible representations, which corresponds to $(T^a)_j^{~i} (T^b)_l^{~k} (P_R)^{jl;i'k'}_{ik;j'l'}$.
We use the short-hand notation $a_j^{~i} \equiv (T^a)_j^{~i}$.
We define the traceless part of the product $a_j^{~i} b_l^{~k}$ as follows.
\eqs{
  \tilde a_j^{~i} \tilde b_l^{~k} & \equiv 
  a_j^{~i} b_l^{~k} 
  - \frac{1}{(N_f^2-4)(N_f^2-1)} \left[ 3N_f \delta_j^{~i} \delta_l^{~k} - (N_f^2+2) \delta_j^{~k} \delta_l^{~i} \right] \mathrm{tr}(ab) \\
  & + \frac{1}{(N_f^2-4)} \left[\delta_j^{~i} (a_l^{~m} b_m^{~k} + a_m^{~k} b_l^{~m}) + \delta_l^{~k} (a_j^{~m} b_m^{~i} + a_m^{~i} b_j^{~m}) \right] \\
  & - \frac{2}{N_f(N_f^2-4)} (\delta_l^{~i} a_m^{~k} b_k^{~m} + \delta_j^{~k} a_l^{~m} b_m^{~i})
  - \frac{N_f^2 - 2}{N_f(N_f^2-4)} (\delta_l^{~i} a_j^{~m} b_m^{~k} + \delta_j^{~k} a_m^{~i} b_l^{~m}) \,.
}
Here, the trace is given by $\mathrm{tr}(ab) = a_j^{~i} b_i^{~j}$.
Using this quantity, we can decompose $a_j^{~i} b_l^{~k}$ into their irreducible parts.
We introduce $(P_R^{ab})_{jl}^{ik}$ as a projection of $a_j^{~i} b_l^{~k}$ into its irreduceble representation $R$
\eqs{
  a_j^{~i} b_l^{~k} = \sum_R (P_R^{ab})_{jl}^{ik}\,,
}
and each representation takes the form
\begin{align}
  (P_S^{ab})_{jl}^{ik} & = \frac{1}{4} (\tilde a_j^{~i} \tilde b_l^{~k} + \tilde a_j^{~k} \tilde b_l^{~i} + \tilde a_l^{~i} \tilde b_k^{~k} + \tilde a_l^{~k} \tilde b_j^{~i}) \,, \\
  (P_Y^{ab})_{jl}^{ik} & = \frac{1}{4} (\tilde a_j^{~i} \tilde b_l^{~k} - \tilde a_j^{~k} \tilde b_l^{~i} - \tilde a_l^{~i} \tilde b_k^{~k} + \tilde a_l^{~k} \tilde b_j^{~i}) \,, \\
  (P_T^{ab})_{jl}^{ik} & = \frac{1}{4} (\tilde a_j^{~i} \tilde b_l^{~k} + \tilde a_j^{~k} \tilde b_l^{~i} - \tilde a_l^{~i} \tilde b_k^{~k} - \tilde a_l^{~k} \tilde b_j^{~i}) \,, \\
  (P_{\overline T}^{ab})_{jl}^{ik} & = \frac{1}{4} (\tilde a_j^{~i} \tilde b_l^{~k} - \tilde a_j^{~k} \tilde b_l^{~i} + \tilde a_l^{~i} \tilde b_k^{~k} - \tilde a_l^{~k} \tilde b_j^{~i}) \,, \\
  (P_{\mathrm{Ad}_S}^{ab})_{jl}^{ik} & = \frac{1}{2(N_f^2-4)} \biggl[ 
    N_f \delta_l^{~i} \{ a,b\}_j^{~k} + N_f \delta_j^{~k} \{ a,b \}_l^{~i} - \delta_j^{~i} \{ a,b\}_l^{~k} - \delta_l^{~k} \{ a,b \}_j^{~i} \nonumber \\
    & \qquad \left. 
    - 4 \delta_l^{~i}\delta_j^{~k} \mathrm{tr}(ab) + \frac{4}{N_f}\delta_j^{~i}\delta_l^{~k} \mathrm{tr}(ab)
   \right] \,, \\
  (P_{\mathrm{Ad}_A}^{ab})_{jl}^{ik} & = \frac{1}{2N_f} \left( 
    \delta_l^{~i} [ a,b ]_j^{~k} - \delta_j^{~k} [ a,b ]_l^{~i} \right) \,, \\
  (P_1^{ab})_{jl}^{ik} & = \frac{1}{N_f^1-1} \left( \delta_l^{~i} \delta_j^{~k} - \frac{1}{N_f} \delta_j^{~i} \delta_l^{~k} \right) \mathrm{tr}(ab) \,.
\end{align}
Here, $\{ a,b\}_j^{~i} \equiv a_j^{~m} b_m^{~i} + b_j^{~m} a_m^{~i}$ and $[ a,b]_j^{~i} \equiv a_j^{~m} b_m^{~i} - b_j^{~m} a_m^{~i}$.
By multiplying $4 c_i^{~j} d_k^{~l}$, we obtain the projection operators with indices of pions, $P_R^{ab;cd}$.
The projection operators are 
\begin{align}
  P_{1}^{ab;cd} & = \frac{1}{N_f^2-1} \delta^{ab} \delta^{cd} \,, \\
  P_{\mathrm{Ad}_A}^{ab;cd} & = \frac{1}{N_f} f^{abe} f^{cde} \,, \\
  P_{\mathrm{Ad}_S}^{ab;cd} & = \frac{N_f}{N_f^2-4} d^{abe} d^{cde} \,, \\
  P_{T}^{ab;cd} & = \frac{1}{4} (\delta^{ac} \delta^{bd} - \delta^{ad} \delta^{bc}) - \frac{1}{2N_f} f^{abe} f^{cde} \nonumber \\
  & \qquad + \frac{i}{8} (f^{ace} d^{bde} + d^{ace} f^{bde} - f^{ade} d^{bce} - d^{ade} f^{bce}) \,, \\
  P_{\overline T}^{ab;cd} & = \frac{1}{4} (\delta^{ac} \delta^{bd} - \delta^{ad} \delta^{bc}) - \frac{1}{2N_f} f^{abe} f^{cde} \nonumber \\
  & \qquad - \frac{i}{8} (f^{ace} d^{bde} + d^{ace} f^{bde} - f^{ade} d^{bce} - d^{ade} f^{bce}) \,, \\
  P_{Y}^{ab;cd} & = \frac{N_f-2}{4N_f} (\delta^{ac} \delta^{bd} + \delta^{ad} \delta^{bc}) + \frac{N_f-2}{2N_f(N_f-1)}\delta^{ab} \delta^{cd} \nonumber \\
  & \qquad - \frac14 (d^{ace} d^{bde} + d^{ade} d^{bce}) + \frac{N_f-4}{4(N_f-2)} d^{abe} d^{cde} \,, \\
  P_{S}^{ab;cd} & = \frac{N_f+2}{4N_f} (\delta^{ac} \delta^{bd} + \delta^{ad} \delta^{bc}) - \frac{N_f+2}{2N_f(N_f+1)}\delta^{ab} \delta^{cd} \nonumber \\
  & \qquad + \frac14 (d^{ace} d^{bde} + d^{ade} d^{bce}) - \frac{N_f+4}{4(N_f+2)} d^{abe} d^{cde} \,.
\end{align}
These projection operators satisfy 
\eqs{
  \sum_R P_R^{ab;cd} = \delta^{ac} \delta^{bd} \,, \qquad
  \sum_{c',d'} P_{R}^{ab;c'd'} P_{R'}^{c'd';cd} = \delta_{R R'} P_{R}^{ab;cd} \,.
  \label{eq:property_proj}
}
To verify the second equality, we use several identities with $d^{abc}$ and $f^{abc}$, such as 
\eqs{
  f^{acd} f^{bcd} & = - N_f \delta^{ab} \,, \qquad 
  d^{acd} d^{bcd} = \frac{N_f^2-4}{N_f} \delta^{ab} \,, \qquad \\
  f^{abe}f^{cde} & = d^{ace}d^{bde} - d^{ade}d^{bce}  + \frac{2}{N_f}(\delta^{ac}\delta^{bd}-\delta^{ad}\delta^{bc}) \,, \\
  f^{abe}d^{cde} & = d^{ace}f^{bde}+ d^{ade}f^{bce} \,, \\
  f^{abe}f^{cde} & = - f^{ace}f^{dbe} -f^{ade}f^{bce} \,, 
  \label{eq:fd2SUN}
}
and 
\eqs{
  \mathrm{Tr}(F^a F^b F^c F^d) 
  & = \delta^{ab}\delta^{cd} + \delta^{ad}\delta^{bc} + \frac{N_f}{4} (d^{abe} d^{cde} - d^{ace} d^{bde} + d^{ade}d^{bce}) \,, \\
  \mathrm{Tr}(F^a F^b D^c D^d) 
  & = \frac{N_f^2-4}{N_f^2} (\delta^{ab}\delta^{cd} - \delta^{ac}\delta^{bd}) + \frac{N_f^2 -8}{4N_f} (d^{abe} d^{cde} - d^{ace} d^{bde}) + \frac{N_f}{4} d^{ade}d^{bce} \,, \\
  \mathrm{Tr}(D^a D^b D^c D^d) 
  & = \frac{N_f^2-4}{N_f^2} (\delta^{ab}\delta^{cd} + \delta^{ad}\delta^{bc}) + \frac{N_f^2-16}{4N_f} (d^{abe} d^{cde} + d^{ade} d^{bce}) + \frac{N_f}{4} d^{ace}d^{bde} \,,
  \label{eq:fd4SUN}
}
where $(F^a)^{bc} = - i f^{abc}$ and $(D^a)^{bc} = d^{abc}$, and $\mathrm{Tr}$ denotes a trace of adjoint indices. 
The first equation in \cref{eq:fd4SUN} is derived from computing $\mathrm{tr}([[[[T^e,T^a],T^b],T^c],T^d]T^e)$ in two different ways: one way is to use the commutation relation, and another way is to reduce the number of generators by use of the Fierz identity and directly compute the traces of four generators. 
More complicated relations would be required for the decomposition of three products of generators, which also requires the dedicated analysis for the resummation of $3 \to 2$ initial states.
The trace of projection operators gives the dimension of the representation, 
\eqs{
  \sum_{a,b,c,d} P_{R}^{ab;cd} P_{R}^{cd;ab} = d_R \,,
}
where $d_R$ refers to the dimensions of the representation $R$ (see \cref{tab:reps} for the dimensions).

%%%%%%%%%%%%%%%%%%%%%%%%%%
\subsection{\texorpdfstring{$SU(N_f)/SO(N_f)$}{SU(N)/SO(N)}}

Let us consider the symmetry breaking $SU(N_f) \to SO(N_f)$.
The number of unbroken generators is $N_f(N_f-1)/2$, and hence the number of broken generators is $(N_f+2)(N_f-1)/2$.
The broken generator satisfies $T^a = (T^a)^T$, and hence the pions transform as a second rank symmetric tensor field. 
The product of the symmetric tensors is decomposed as follows. 
\eqs{
  s \otimes s
  = 1_S 
  \oplus \mathrm{Ad}_A \oplus s_S
  \oplus T_A \oplus Y_S \oplus S_S \,.
}
Here, the subscripts $S$ and $A$ denote symmetrization and anti-symmetrization of indices of symmetric representations in the left-hand side.
\cref{tab:repsSO} shows the representations and their dimensions.
As seen in the table, several representations vanish for $N_f \leq 3$.
For $N_f = 3$, $Y_S$ does not appear and the decomposition is given by
\eqs{
  \mathbf{5} \otimes \mathbf{5}
  = \mathbf{1}_S \oplus \mathbf{3}_A \oplus \mathbf{5}_S \oplus \mathbf{7}_S \oplus \mathbf{9}_S \,.
}

\begin{table}[t]
  \centering
	\caption{
    Representations in $SO(N_f)$ and their dimensions
	}
	\label{tab:repsSO}
  \renewcommand\arraystretch{2}
  \begin{tabular}{|c|c|}
    \hline
    Reps. & Dims. \\ \hline
    $s$ & $\dfrac12 (N_f-1) (N_f+2)$ \\
    $\mathrm{Ad}$ & $\dfrac12 N_f (N_f-1)$ \\
    $Y$ & $\dfrac{1}{12} (N_f-3) N_f (N_f+1) (N_f+2)$ \\
    $S$ & $\dfrac{1}{24} (N_f^2-1) N_f (N_f+6)$ \\
    $T$ & $\dfrac18 (N_f^2-1) (N_f+4) (N_f-2)$ \\ \hline
  \end{tabular}
\end{table}

It is convenient to make the upper indices of the broken generator lower since the broken generators are symmetric: $(T^a)_{ij} \equiv (T^a)_{i}^{~j}$.
The Fierz identity for the broken generator is given by 
\eqs{
  (T^a)_{ij} (T^a)_{kl} = \frac14 \left( \delta_{ik} \delta_{jl} + \delta_{il} \delta_{jk} - \frac{2}{N_f} \delta_{ij} \delta_{kl} \right) \,.
  \label{eq:FierzSO}
}
Here, the summation over $a$ runs only over the broken generators.
We construct the projection operators for the product of broken generators as with the $SU(N_f)$. 
The projection operators with the labels of the broken generators are defined by 
\eqs{
  P_R^{ab;cd} = 4 (T^a)_{ij} (T^b)_{kl} (P_R)_{ijkl;i'j'k'l'} (T^c)_{i'j'} (T^d)_{k'l'} \,.
}
Now, we decompose the product $(T^a)_{ij} (T^b)_{kl}$ into its irreducible representations, which corresponds to $(T^a)_{ij} (T^b)_{kl} (P_R)_{ijkl;i'j'k'l'}$.
We use the short-hand notation $a_{ij} \equiv (T^a)_{ij}$.
We define the traceless part of the product $a_{ij} b_{kl}$ as follows.
\eqs{
  & \tilde a_{ij} \tilde b_{kl} \equiv 
  a_{ij} b_{kl} \\
  & \quad - \frac{1}{(N_f^2-4)(N_f-1)(N_f+4)} \left[ 2 (3N_f+2) \delta_{ij} \delta_{kl} - (N_f^2+4) (\delta_{il} \delta_{jk} + \delta_{ik} \delta_{jl})\right] \mathrm{tr}(ab) \\
  & \quad + \frac{2}{(N_f+4)(N_f-2)} (\delta_{ij} a_{km} b_{ml} + \delta_{ij} a_{lm} b_{mk} + \delta_{kl} a_{im} b_{mj} + \delta_{kl} a_{jm} b_{mi}) \\
  & \quad - \frac{4}{(N_f^2-4)(N_f+4)} (\delta_{il} a_{km} b_{mj} + \delta_{jl} a_{km} b_{mi} + \delta_{ik} a_{lm} b_{mj} + \delta_{jk} a_{lm} b_{mi}) \\
  & \quad - \frac{N_f^2 + 2 N_f - 4}{(N_f^2 -4)(N_f+4)} (\delta_{il} a_{jm} b_{mk} + \delta_{jl} a_{im} b_{mk} + \delta_{ik} a_{jm} b_{ml} + \delta_{jk} a_{im} b_{ml}) \,.
}
The trace is defined by $\mathrm{tr}(ab) \equiv a_{ij} b_{ji}$.
Using this quantity, we can decompose $a_{ij} b_{kl}$ into their irreducible parts.
We introduce $(P_R^{ab})_{ijkl}$ as a projection of $a_{ij} b_{kl}$ into its irreducible representation $R$
\eqs{
  a_{ij} b_{kl} = \sum_R (P_R^{ab})_{ijkl}\,,
}
and each representation takes the form
\begin{align}
  (P_S^{ab})_{ijkl} & = \frac{1}{6} (\tilde a_{ij} \tilde b_{kl} + \tilde a_{kj} \tilde b_{il} + \tilde a_{ki} \tilde b_{jl} + \tilde a_{lj} \tilde b_{ik} + \tilde a_{il} \tilde b_{kj} + \tilde a_{kl} \tilde b_{ij}) \,, \\
  (P_Y^{ab})_{ijkl} & = \frac{1}{6} (2 \tilde a_{ij} \tilde b_{kl} - \tilde a_{kj} \tilde b_{il} - \tilde a_{ki} \tilde b_{jl} - \tilde a_{lj} \tilde b_{ik} - \tilde a_{il} \tilde b_{kj} + 2 \tilde a_{kl} \tilde b_{ij}) \,, \\
  (P_T^{ab})_{ijkl} & = \frac{1}{2} (\tilde a_{ij} \tilde b_{kl} - \tilde a_{kl} \tilde b_{ij}) \,, \\
  (P_s^{ab})_{ijkl} & = - \frac{2}{(N_f+4)(N_f-2)} \biggl[ 
    \delta_{ij} \{ a,b\}_{kl} + \delta_{kl} \{ a,b \}_{ij} \nonumber \\
    & \qquad 
    - \frac{N_f}{4} \left( \delta_{il} \{ a,b \}_{jk} + \delta_{ik} \{ a,b \}_{jl} + \delta_{jl} \{ a,b \}_{ik} + \delta_{jk} \{ a,b \}_{il}\right) \nonumber \\
    & \qquad \left. 
    - \frac{4}{N_f} \delta_{ij}\delta_{kl} \mathrm{tr}(ab) + (\delta_{il} \delta_{jk} + \delta_{ik}\delta_{jl}) \mathrm{tr}(ab)
   \right] \,, \\
  (P_\mathrm{Ad}^{ab})_{ijkl} & = \frac{1}{2(N_f+2)} \left( 
  \delta_{il} [ a,b ]_{jk} + \delta_{ik} [ a,b ]_{jl} + \delta_{jl} [ a,b ]_{ik} + \delta_{jk} [ a,b ]_{il} \right) \,, \\
  (P_1^{ab})_{ijkl} & = \frac{1}{(N_f-1)(N_f+2)} \left[ \delta_{il} \delta_{jk} + \delta_{ik} \delta_{jl} - \frac{2}{N_f} \delta_{ij} \delta_{kl} \right] \mathrm{tr}(ab) \,.
\end{align}
By multiplying $4 c_{ij} d_{kl}$, we obtain the projection operators with indices of pions, $P_R^{ab;cd}$.
Since the broken generators are parts of $SU(N_f)$ generators, we may write the projection operators in terms of the group-theoretical constants of $SU(N_f)$, $f_{abc}$ and $d_{abc}$, as in the previous subsection.
Hence, the commutation and anticommutation relations of the broken generators are given by
\eqs{
  [T^a, T^b] = i f^{ab \bar c} T^{\bar c} \,, \qquad 
  \{T^a, T^b\} = d^{ab \bar c} T^{\bar c} + \frac{1}{N_f} \delta^{ab} \,,
}
where the indices with bars run over all generators of $SU(N_f)$, but those without bars denote the broken generators.
We find $f^{ab \bar c}$ is non-zero only when $\bar c$ runs over the unbroken generators, while $d^{ab \bar c}$ is non-zero only when $\bar c$ runs over the broken generators, which are known as Cartan decomposition.

We find the projection operators $P_R^{ab;cd}$ in $SU(N_f)/SO(N_f)$ as follows.
\begin{align}
  P_{1}^{ab;cd} & = \frac{2}{(N_f-1)(N_f+2)} \delta^{ab} \delta^{cd} \,, \\
  P_{\mathrm{Ad}}^{ab;cd} & = \frac{2}{N_f+2} f^{ab\bar{e}} f^{cd\bar{e}} \,, \\
  P_{s}^{ab;cd} & = \frac{2N_f}{(N_f+4)(N_f-2)} d^{ab\bar{e}} d^{cd\bar{e}} \,, \\
  P_{T}^{ab;cd} & = \frac{1}{2} (\delta^{ac} \delta^{bd} - \delta^{ad} \delta^{bc}) - \frac{2}{N_f+2} f^{ab\bar{e}} f^{cd\bar{e}} \,, \\
  P_{Y}^{ab;cd} & = \frac{N_f-2}{3N_f} (\delta^{ac} \delta^{bd} + \delta^{ad} \delta^{bc}) + \frac{2(N_f-2)}{3N_f(N_f-1)} \delta^{ab} \delta^{cd} \nonumber \\
  & \qquad - \frac13 (d^{ac\bar{e}} d^{bd\bar{e}} + d^{ad\bar{e}} d^{bc\bar{e}}) + \frac{N_f-4}{3(N_f-2)} d^{ab\bar{e}} d^{cd\bar{e}} \,, \\
  P_{S}^{ab;cd} & = \frac{N_f+4}{6N_f} (\delta^{ac} \delta^{bd} + \delta^{ad} \delta^{bc}) - \frac{2(N_f+4)}{3N_f(N_f+2)} \delta^{ab} \delta^{cd} \nonumber \\
  & \qquad + \frac13 (d^{ac\bar{e}} d^{bd\bar{e}} + d^{ad\bar{e}} d^{bc\bar{e}}) 
  - \frac{N_f+8}{3(N_f+4)} d^{ab\bar{e}} d^{cd\bar{e}} \,.
\end{align}
These are consistent with the projection operators given in Ref.~\cite{Bijnens:2011fm}.

The trace of projection operators gives the dimension of the representation as with the previous subsection, 
\eqs{
  \sum_{a,b,c,d} P_{R}^{ab;cd} P_{R}^{cd;ab} = d_R \,,
}
where $d_R$ refers to the dimensions of the representation $R$ (see \cref{tab:repsSO} for the dimensions).
We confirm properties of the projection operators by taking the trace of the projection operators:
\eqs{
  \sum_R \sum_{a,b} P_R^{ab;ab} = \frac14 (N_f-1)^2(N_f+2)^2 \,, \qquad
  \sum_{a,b,c,d} P_{R}^{ab;cd} P_{R'}^{ab;cd} = \delta_{R R'} d_{R} \,.
}
Here, we use the Fierz identity for the broken generators [see \cref{eq:FierzSO}] instead of \cref{eq:fd2SUN} modified to apply it for the broken generators.

%%%%%%%%%%%%%%%%%%%%%%%%%%
\subsection{\texorpdfstring{$SU(N_f)/USp(N_f)$}{SU(N)/USp(N)}}

Let us consider the symmetry breaking $SU(N_f) \to USp(N_f)$.
The number of unbroken generators is $N_f(N_f+1)/2$, and hence the number of broken generators is $(N_f-2)(N_f+1)/2$.
The broken generator satisfies $T^a J = J (T^a)^T = - (T^a J)^T$ with $J$ being the symplectic metric, and hence the pions transform as a second rank antisymmetric tensor field. 
The product of the antisymmetric tensors is decomposed as follows. 
\eqs{
  a \otimes a
  = 1_S 
  \oplus a_S \oplus \mathrm{Ad}_A
  \oplus T_A \oplus Y_S \oplus S_S \,.
}
Here, the subscripts $S$ and $A$ denote symmetrization and anti-symmetrization of indices of pions (or $\pi^a T^a$) in the left-hand side.
\cref{tab:repsSp} shows the representations and their dimensions.
As seen in the table, several representations vanish for $N_f \leq 6$.
For $N_f = 4$, $a_S$, $T_A$, and $S_S$ do not appear and the decomposition is given by
\eqs{
  \mathbf{5} \otimes \mathbf{5}
  = \mathbf{1}_S \oplus \mathbf{10}_A \oplus \mathbf{14}_S \,,
}
and for $N_f = 6$, $S_S$ does not appear and the decomposition is given by
\eqs{
  \mathbf{14} \otimes \mathbf{14}
  = \mathbf{1}_S \oplus \mathbf{14}_S \oplus \mathbf{21}_A \oplus \mathbf{70}_A \oplus \mathbf{90}_S \,.
}

\begin{table}[t]
  \centering
	\caption{Representations in $USp(N_f)$ and their dimensions
	}
	\label{tab:repsSp}
  \renewcommand\arraystretch{2}
  \begin{tabular}{|c|c|}
    \hline
    Reps. & Dims. \\ \hline
    $a$ & $\dfrac12 (N_f-2) (N_f+1)$ \\
    $\mathrm{Ad}$ & $\dfrac12 N_f (N_f+1)$ \\
    $Y$ & $\dfrac{1}{12} (N_f-2) (N_f-1) N_f (N_f+3)$ \\
    $S$ & $\dfrac{1}{24} (N_f^2 -1) N_f (N_f-6)$ \\
    $T$ & $\dfrac{1}{8} (N_f^2 -1) (N_f-4) (N_f+2)$ \\ \hline
  \end{tabular}
\end{table}

We construct the projection operators for the product of broken generators as with the previous subsections.
We define the generator with the symplectic metric as $(\tau^a)_{ij} \equiv (T^a)_i^{j'} J_{jj'}$, which satisfies $(\tau^a)^T = - \tau^a$.
The Fierz identity for the broken generator is given by 
\eqs{
  (\tau^a)_{ij} (\tau^a)_{kl} = \frac14 \left( J_{ik} J_{jl} - J_{il} J_{jk} - \frac{2}{N_f} J_{ij} J_{kl} \right) \,.
  \label{eq:FierzSp}
}

The projection operators with the labels of the broken generators are defined by 
\eqs{
  P_R^{ab;cd} = 4 (\tau^a)_{ij} (\tau^b)_{kl} (P_R)^{ijkl;i'j'k'l'} (\tau^c)_{i'j'} (\tau^d)_{k'l'} \,.
}
Now, we decompose the product $(\tau^a)_{ij} (\tau^b)_{kl}$ into its irreducible representations as with previous subsection.
We use the short-hand notation $a_{ij} \equiv (\tau^a)_{ij}$.
We define the traceless part of the product $a_{ij} b_{kl}$ as follows.
\eqs{
  & \tilde a_{ij} \tilde b_{kl} \equiv 
  a_{ij} b_{kl} \\
  & \quad - \frac{1}{(N_f^2 -4)(N_f+1)(N_f-4)} \left[ 2 (3N_f+2) J_{ij} J_{kl} + (N_f^2 +4) (J_{il} J_{jk} - J_{ik} J_{jl})\right] \mathrm{tr}(ab) \\
  & \quad + \frac{2}{(N_f-4)(N_f+2)} (J_{ij} J^{mn} a_{km} b_{nl} + J_{ij} J^{mn} a_{lm} b_{nk} + J_{kl} J^{mn} a_{im} b_{nj} + J_{kl} J^{mn} a_{jm} b_{ni}) \\
  & \quad - \frac{4}{(N_f^2 -4)(N_f-4)} (J_{il} J^{mn} a_{km} b_{nj} - J_{jl} J^{mn} a_{km} b_{ni} - J_{ik} J^{mn} a_{lm} b_{nj} + J_{jk} J^{mn} a_{lm} b_{ni}) \\
  & \quad + \frac{N_f^2  - 2 N_f - 4}{(N_f^2 -4)(N_f-4)} (J_{il} J^{mn} a_{jm} b_{nk} - J_{jl} J^{mn} a_{im} b_{nk} - J_{ik} J^{mn} a_{jm} b_{nl} + J_{jk} J^{mn} a_{im} b_{nl}) \,.
}
Here, the invariant tensor is $J_{ij}$, but not $\delta_{ij}$, and the trace is defined by $\mathrm{tr}(ab) \equiv J^{ik} J^{jl} a_{ij} b_{kl}$.
This trace coincides with the trace of $SU(N_f)$ generators $T^a$ as follows:
\eqs{
  \mathrm{tr}(ab) 
  & = J^{ik} J^{jl} (\tau^a)_{ij} (\tau^b)_{kl} 
  = - J^{ik} J^{jl} J_{jj'} J_{kk'} (T^a)_i^{j'} (T^b)_l^{k'} \\
  & = \delta^i_{k'} \delta^l_{j'} (T^a)_i^{j'} (T^b)_l^{k'}
  = \mathrm{tr}(T^a T^b) \,.
}
Using the traceless part, we can decompose $a_{ij} b_{kl}$ into their irreducible parts.
We introduce $(P_R^{ab})_{ijkl}$ as a projection of $a_{ij} b_{kl}$ into its irreducible representation $R$,
\eqs{
  a_{ij} b_{kl} = \sum_R (P_R^{ab})_{ijkl}\,,
}
and each representation takes the form
\begin{align}
  (P_S^{ab})_{ijkl} & = \frac{1}{6} (\tilde a_{ij} \tilde b_{kl} - \tilde a_{ik} \tilde b_{jl} + \tilde a_{il} \tilde b_{jk} + \tilde a_{jk} \tilde b_{il} - \tilde a_{jl} \tilde b_{ik} + \tilde a_{kl} \tilde b_{ij}) \,, \\
  (P_Y^{ab})_{ijkl} & = \frac{1}{6} (2 \tilde a_{ij} \tilde b_{kl} + \tilde a_{ik} \tilde b_{jl} - \tilde a_{il} \tilde b_{jk} - \tilde a_{jk} \tilde b_{il} + \tilde a_{jl} \tilde b_{ik} + 2 \tilde a_{kl} \tilde b_{ij}) \,, \\
  (P_T^{ab})_{ijkl} & = \frac{1}{2} (\tilde a_{ij} \tilde b_{kl} - \tilde a_{kl} \tilde b_{ij}) \,, \\
  (P_a^{ab})_{ijkl} & = - \frac{2}{(N_f-4)(N_f+2)} \biggl[ 
    J_{ij} \{ a,b\}_{kl} + J_{kl} \{ a,b \}_{ij} \nonumber \\
    & \qquad 
    - \frac{N_f}{4} \left( J_{ik} \{ a,b \}_{jl} + J_{jl} \{ a,b \}_{ik} - J_{il} \{ a,b \}_{jk} - J_{jk} \{ a,b \}_{il}\right) \nonumber \\
    & \qquad \left. 
    + \frac{4}{N_f} J_{ij}J_{kl} \mathrm{tr}(ab) - (J_{il} J_{jk} - J_{ik}J_{jl}) \mathrm{tr}(ab)
   \right] \,, \\
  (P_\mathrm{Ad}^{ab})_{ijkl} & = \frac{1}{2(N_f-2)} \left( 
  J_{ik} [ a,b ]_{jl} - J_{jk} [ a,b ]_{il} - J_{il} [ a,b ]_{jk} + J_{jl} [ a,b ]_{ik} \right) \,, \\
  (P_1^{ab})_{ijkl} & = - \frac{1}{(N_f+1)(N_f-2)} \left[ J_{il} J_{jk} - J_{ik} J_{jl} - \frac{2}{N_f} J_{ij} J_{kl} \right] \mathrm{tr}(ab) \,.
\end{align}

We obtain the projection operators in terms of the pion indices by multiplying $4c^{ij} d^{kl}$.
Similarly to the previous subsection, the broken generators are parts of the $SU(N_f)$ generators, and hence we may write the projection operators in terms of the group-theoretical constants for the $SU(N_f)$ generators.
\begin{align}
  P_{1}^{ab;cd} & = \frac{2}{(N_f-2)(N_f+1)} \delta^{ab} \delta^{cd} \,, \\
  P_{\mathrm{Ad}}^{ab;cd} & = \frac{2}{N_f-2} f^{ab\bar{e}} f^{cd\bar{e}} \,, \\
  P_{a}^{ab;cd} & = \frac{2N_f}{(N_f-4)(N_f+2)} d^{abe\bar{}} d^{cd\bar{e}} \,, \\
  P_{T}^{ab;cd} & = \frac{1}{2} (\delta^{ac} \delta^{bd} - \delta^{ad} \delta^{bc}) - \frac{2}{N_f-2} f^{ab\bar{e}} f^{cd\bar{e}} \,, \\
  P_{Y}^{ab;cd} & = \frac{N_f+2}{3N_f} (\delta^{ac} \delta^{bd} + \delta^{ad} \delta^{bc}) - \frac{2(N_f+2)}{3N_f(N_f+1)} \delta^{ab} \delta^{cd} \nonumber \\
  & \qquad + \frac13 (d^{ac\bar{e}} d^{bd\bar{e}} + d^{ad\bar{e}} d^{bc\bar{e}}) - \frac{N_f+4}{3(N_f+2)} d^{ab\bar{e}} d^{cd\bar{e}} \,, \\
  P_{S}^{ab;cd} & = \frac{N_f-4}{6N_f} (\delta^{ac} \delta^{bd} + \delta^{ad} \delta^{bc}) + \frac{2(N_f-4)}{3N_f(N_f-2)} \delta^{ab} \delta^{cd} \nonumber \\
  & \qquad - \frac13 (d^{ac\bar{e}} d^{bd\bar{e}} + d^{ad\bar{e}} d^{bc\bar{e}}) 
  + \frac{N_f-8}{3(N_f-4)} d^{ab\bar{e}} d^{cd\bar{e}} \,.
\end{align}
Here, the indices without bars denote the labels of broken generator, and the indices with bars run over all $SU(N_f)$ generators.
These are consistent with the projection operators given in Ref.~\cite{Bijnens:2011fm}.

Again, the trace of projection operators gives the dimension of the representation, 
\eqs{
  \sum_{a,b,c,d} P_{R}^{ab;cd} P_{R}^{cd;ab} = d_R \,,
}
where $d_R$ refers to the dimensions of the representation $R$ (see \cref{tab:repsSp} for the dimensions).
We confirm properties of the projection operators by taking the trace of the projection operators:
\eqs{
  \sum_R \sum_{a,b} P_R^{ab;ab} = \frac14 (N_f-2)^2(N_f+1)^2 \,, \qquad
  \sum_{a,b,c,d} P_{R}^{ab;cd} P_{R'}^{ab;cd} = \delta_{R R'} d_{R} \,.
}
Here, as we discussed in the previous section, we use the Fierz identity for the broken generators [see \cref{eq:FierzSp}] instead of \cref{eq:fd2SUN}.

\subsection{Partial-wave amplitudes in 2-to-2 vertices \label{app:coefficients}}

We summarize the tree-level partial-wave amplitudes $T_\ell^R$, which are defined in \cref{eq:pwdecomp}, for each symmetry breaking. 
The symmetric representations only have the $s$-wave amplitude ($T_{\ell=1}^R = 0$ for a symmetric representation $R$), while the antisymmetric representations only have the $p$-wave amplitude ($T_{\ell=0}^R = 0$ for an antisymmetric representation $R$).
For $G = SU(N_f) \times SU(N_f)$ and $H = SU(N_f)$, the partial-wave amplitudes are 
\eqs{
  T_{\ell=0}^{Y} & = \frac{1}{32\pi} \frac{s - 2 m_\pi^2}{f_\pi^2} \,, \qquad 
  T_{\ell=0}^{S} = - \frac{1}{32\pi} \frac{s - 2 m_\pi^2}{f_\pi^2} \,, \\
  T_{\ell=0}^{\mathrm{Ad}_S} & = \frac{1}{32\pi} \frac{1}{f_\pi^2}\left( \frac{N_f}{2} s - \frac{4}{N_f} m_\pi^2 \right) \,, \qquad 
  T_{\ell=0}^{1} = \frac{1}{32\pi} \frac{1}{f_\pi^2}\left( N_f s - \frac{2}{N_f} m_\pi^2 \right) \,,
   \\
  T_{\ell=1}^{\mathrm{Ad}_A} & = \frac{1}{96\pi} \frac{2N_f}{f_\pi^2} \frac{s \beta^2}{4} \,, \qquad 
  T_{\ell=1}^{I} = 0 \quad (I = T \,, \overline T) \,.
  \label{eq:coeff_SU}
}
For $G = SU(N_f)$ and $H = SO(N_f)$, the partial-wave amplitudes are 
\eqs{
  T_{\ell=0}^{Y} & = \frac{1}{32\pi} \frac{s - 2 m_\pi^2}{2 f_\pi^2} \,, \qquad 
  T_{\ell=0}^{S} = - \frac{1}{32\pi} \frac{s - 2 m_\pi^2}{f_\pi^2} \,, \\
  T_{\ell=0}^{s} & = \frac{1}{32\pi} \frac{1}{f_\pi^2}\left( \frac{N_f}{4} s + \frac{N_f-4}{N_f} m_\pi^2 \right) \,, \qquad
  T_{\ell=0}^{1} = \frac{1}{32\pi} \frac{1}{f_\pi^2}\left( \frac{N_f}{2} s + \frac{N_f-2}{N_f} m_\pi^2 \right) \,, \\
  T_{\ell=1}^{\mathrm{Ad}} & = \frac{1}{96\pi} \frac{N_f+2}{f_\pi^2} \frac{s \beta^2}{4} \,, \qquad 
  T_{\ell=1}^{T} = 0 \,.
  \label{eq:coeff_SO}
}
For $G = SU(N_f)$ and $H = USp(N_f)$, the partial-wave amplitudes are 
\eqs{
  T_{\ell=0}^{Y} & = - \frac{1}{32\pi} \frac{s - 2 m_\pi^2}{2 f_\pi^2} \,, \qquad 
  T_{\ell=0}^{S} = \frac{1}{32\pi} \frac{s - 2 m_\pi^2}{f_\pi^2} \,, \\
  T_{\ell=0}^{a} & = \frac{1}{32\pi} \frac{1}{f_\pi^2}\left( \frac{N_f}{4} s - \frac{N_f+4}{N_f} m_\pi^2 \right) \,, \qquad
  T_{\ell=0}^{1} = \frac{1}{32\pi} \frac{1}{f_\pi^2}\left( \frac{N_f}{2}  s - \frac{N_f+2}{N_f} m_\pi^2 \right) \,, \\
  T_{\ell=1}^\mathrm{Ad} & = \frac{1}{96\pi} \frac{N_f-2}{f_\pi^2} \frac{s \beta^2}{4} \,, \qquad 
  T_{\ell=1}^{T} = 0 \,.
  \label{eq:coeff_USp}
}
We note that the projection operators for the higher-dimensional antisymmetric representation do not appear in this vertex function.

\subsection{Group Factors}

The group-theoretical constants used in four-pion interactions [see \cref{eq:4pi}] are defined by
\eqs{
  r_{abcd} & \equiv \frac23 \mathrm{Tr} \left( [T^a,T^c][T^b,T^d] + [T^a,T^d][T^b,T^c] \right) 
  = - \frac13 (f_{ac\bar e}f_{bd\bar e} + f_{ad\bar e} f_{bc\bar e})\,, \\
  c_{abcd} & \equiv \frac13 \mathrm{Tr} \left( T^{\{a} T^b T^c T^{d\}} \right) \\
  & = \frac{1}{3} (d_{ab\bar e}d_{cd\bar e} + d_{ac\bar e}d_{bd\bar e} + d_{ad\bar e}d_{bc\bar e}) + \frac{2}{3 N_f} (\delta_{ab}\delta_{cd} + \delta_{ac}\delta_{bd} + \delta_{ad}\delta_{bc}) \,. 
  \label{eq:r_and_c}
}
Here, the brace in the definition of $c_{abcd}$ denotes the sum of the symmetric permutations. 
Again, the indices without bars run only over the broken generators, while the indices with bars run over all $SU(N_f)$ generators.
The self-scattering cross section of dark pions depends on a group-theoretical factor $a^2$.
\eqs{
  \sigma_\mathrm{scatter} = \frac{m_\pi^2}{32 \pi f_\pi^4} \frac{a^2}{N_\pi^2} \,.
} 
Here, $a^2$ is given by the partial-wave amplitudes in the non-relativistic limit of the two-body system as follows. 
\begin{align}
  SU(N_f) &: \qquad
  a^2 = \frac{1}{2^4} \frac{(N_f^2 -1)(3 N_f^4 -2 N_f^2 + 6)}{2 N_f^2} \,, \\
  SO(N_f) &: \qquad
  a^2 = \frac{1}{2^4} \frac{(N_f -1)(N_f +2)(3 N_f^4 + 7 N_f^3 -2 N_f^2 -12 N_f + 24)}{16N_f^2} \,, \\
  USp(N_f) &: \qquad
  a^2 = \frac{1}{2^4} \frac{(N_f +1)(N_f -2)(3 N_f^4 - 7 N_f^3 -2 N_f^2 +12 N_f + 24)}{16N_f^2} \,. 
\end{align}
These are consistent with the results in the original paper \cite{Hochberg:2014kqa}, which are directly computed from the group-theoretical constants. 
\eqs{
  a^2 = \frac{1}{64} \sum_{abcd}(c_{abcd}^2 + 16 r_{abcd}^2) \,.
}

Last but not least, we show the coefficients $t_R^2$, which are used in \cref{eq:thermal_averaged_3-to-2}, for different symmetries. 
We compute $t_R^2$ by using the Fierz identity for the generators instead of computing the group-theoretical constants, $d$ and $f$, since all pion indices are contracted. 
These are given by
\begin{align}
  SU(N_f) &:&
  t_{\mathrm{Ad}}^2 & = \frac{1}{2^5} \frac{N_f (N_f^2 -1)(N_f^2 -4)}{48} \,, &
  t_{T,\overline T}^2 & = \frac{1}{2^5} \frac{N_f (N_f^2 -1)(N_f^2 -4)}{96} \,, \\
  SO(N_f) &:& 
  t_{\mathrm{Ad}}^2 & = \frac{1}{2^5} \frac{N_f^2 (N_f^2 -1)(N_f -2)}{768} \,, &
  t_{T}^2 & = \frac{1}{2^5} \frac{N_f (N_f+4) (N_f^2 -1)(N_f -2)}{768} \,, \\
  USp(N_f) &:& 
  t_{\mathrm{Ad}}^2 & = \frac{1}{2^5} \frac{N_f^2 (N_f^2 -1)(N_f +2)}{768} \,, &
  t_{T}^2 & = \frac{1}{2^5} \frac{N_f (N_f-4) (N_f^2 -1)(N_f +2)}{768} \,, 
\end{align}
and others not listed here are zero.
The sum of them is consistent with the values given by \cite{Hochberg:2014kqa}.
\begin{align}
  SU(N_f) &: \qquad
  t^2 = \frac{1}{4^5} \frac{4 N_f (N_f^2 -1)(N_f^2 -4)}{3}  \,, \\
  SO(N_f) &: \qquad
  t^2 = \frac{1}{4^5} \frac{N_f (N_f^2 -1)(N_f^2 -4)}{12} \,, \\
  USp(N_f) &: \qquad
  t^2 = \frac{1}{4^5} \frac{N_f (N_f^2 -1)(N_f^2 -4)}{12} \,.
\end{align}

\bibliographystyle{utphys}
\bibliography{ref}

\end{document}